\documentclass[]{article}

\usepackage{amsfonts,amsmath,amssymb}
\usepackage{amsthm}
\usepackage[cp866]{inputenc}
\usepackage[english]{babel}

\allowdisplaybreaks \theoremstyle{definition}
\newtheorem{Definition}{Definition}
\newtheorem{Remark}{Remark}
\newtheorem{Remarks}{Remarks}

\theoremstyle{plain}
\newtheorem{Theorem}{Theorem}
\newtheorem{Corollary}{Corollary}
\newtheorem{Proposition}{Proposition}

\newtheorem{Lemma}{Lemma}

\newcommand \rat {{\mathbb Q}}
\newcommand \intt {{\mathbb Z}}
\newcommand \real {{\mathbb R}}

\newcommand \nat {{\mathbb N}}
\newcommand \gs {\geqslant}
\newcommand \ls {\leqslant}

\newcommand \quot {\text {\rm quot}}
\newcommand \rest {\text {\rm rest}}
\newcommand \degr {\text {\rm deg}_{\text {\rm R}}}
\newcommand \sseq {\text {\rm sseq}}

\newcommand \res {\text {\rm res}}

\newcommand \ra {{\mathbb R}_{\text {\rm alg}}}
\newcommand \rc {{\mathbb R}_{\text {\rm c}}}
\newcommand \rp {{\mathbb R}_{\text {\rm p}}}

\newcommand \co {{\mathbb C}}

\newcommand \dv {\,|\,}

\newcommand \sq {\subseteq}
\newcommand \sign {{\text {\rm sign}}}
\newcommand \spec {{\text {\rm spec}}}

\newcommand \bft {{\mathbf t}}
\newcommand \cau {\mathbf{Ca}}

\newcommand{\rng}{\mathit{rng}}

\newcommand{\calN}{\mathcal{N}}

\newcommand{\IR}{\mathbb{R}}
\newcommand{\prs}{\rm pras}
\newcommand{\pr}{\rm pra}
\newcommand{\prr}{\rm pro}
\newcommand{\com}{\rm co}

\newcommand{\cs}{\rm cs}
\title{Primitive Recursive Ordered Fields
and Some Applications\footnote{This is an extended version of the conference paper \cite{ss21} which contains new results,  detailed proofs of some results which were only sketched in the conference version, and precise formulations of  auxiliary facts.}}

\author{Victor  Selivanov\thanks{The work of V. Selivanov is supported by Mathematical Center in Akademgorodok under agreement No. 075-15-2019-1613 with the Ministry of Science and Higher Education of the Russian Federation.}
 \\A.P. Ershov Institute of
Informatics Systems\\ and
S.L. Sobolev Institute of Mathematics
\\{\tt vseliv@iis.nsk.su}
\\ and
\\Svetlana Selivanova\thanks{The work of S. Selivanova is partially supported by RFBR-JSPS Grant 20-51-50001, by the National Research Foundation of Korea
(grant 2017R1E1A1A03071032), 
by the International Research \& Development Program of
the Korean Ministry of Science and ICT (grant 2016K1A3A7A03950702),
and by the NRF Brain Pool program (grant 2019H1D3A2A02102240).}
\\Korea Advanced Institute of Science and Technology \\{\tt sseliv@kaist.ac.kr}
}

\begin{document}
\large
\date{}

 \maketitle

\begin{abstract}
We  establish  primitive recursive versions of some known facts about computable ordered fields of reals and computable reals, and then apply them to proving primitive recursiveness of some natural problems in linear algebra and analysis. In particular, we find a partial primitive recursive analogue of Ershov-Madison's theorem  about real closures of computable ordered fields, relate the corresponding fields to the primitive recursive reals, give  sufficient conditions for primitive recursive root-finding,   computing normal forms of matrices, and  computing solution operators of some linear systems of PDE.

\end{abstract}

\section{Introduction}\label{in}

In \cite{ss17,ss17a},  computable ordered fields of reals  were related to the field of computable reals and were used to prove computability of some problems in algebra and analysis (notably,  spectral problems for symmetric matrices and computing solutions of symmetric hyperbolic systems of partial differential equations (PDEs) uniformly on matrix coefficients) in the rigorous sense of computable analysis \cite{wei}. The proposed sufficient conditions for computability are very broad  but  do not yield any complexity upper bounds because they  use algorithms based on unbounded search through countable sets. We note that the situation here is rather subtle, e.g. the spectral decomposition of a real symmetric $2\times2$-matrix is not computable \cite{zb04} (as a multi-valued function on the reals) but it becomes computable (even for $n\times n$-matrices uniformly on $n$) if matrix coefficients range over any fixed computable real closed field of reals.

In \cite{ss18}, the PTIME-presentability of the ordered field $\ra$ of algebraic reals and PTIME-computability of some problems on algebraic numbers established in \cite{as18} were applied to find non-trivial upper complexity bounds for the aforementioned problems in algebra and analysis. A weak point here is that this approach applies only to problems with coefficients in $\ra$ because $\ra$ is currently the only known PTIME-presentable real closed ordered field.

Another weak point is that  complexity classes like PTIME or PSPACE are often not closed under important constructions.
E.g. the spectral decomposition of an algebraic symmetric $n\times n$-matrix is PTIME-computable for any fixed $n$, but not uniformly on $n$. The same holds for the problem of root-finding for polynomials in $\ra[x]$ \cite{as18}. For differential equations, PTIME is in many cases preserved for analytic/polynomial initial data: \cite{bgp11, kst18} for ordinary differential equations (ODEs), \cite{KSZ19} for PDEs. However, for more general functional classes the situation is different: solving ODEs with a Lipshitz continuous PTIME computable right-hand part is PSPACE-complete \cite{kaw10}. 
Computing the solutions of the Dirichlet problem for the Poisson equation \cite{KSZ17} and periodic boundary value problem for the heat equation  \cite{KPSZ21}, is $\#$P$_1$-complete; according to \cite{KPSZ21,KSZ19}, for a large class of linear evolutionary PDEs, the difference scheme method is in general in PSPACE, for some particular cases in $\#$P, when applied to fixed real PTIME computable initial data.
 
Thus, it seems reasonable to investigate properties of the aforementioned problems for natural complexity classes in between PTIME and COMPUTABLE, to obtain better closure properties and  efficient solutions of wider classes of problems. An obvious candidate here is the class PR of  primitively recursive functions having a prominent role in computability theory and proof theory.

Recently, there was a renewed interest in primitively recursive (PR) structures (see e.g. \cite{bd19} and references therein) which are recognized as a principal model for an emerging new paradigm of computability --- the so called online computability (see e.g. \cite{br98}). PR-solvability of a problem yields a solution algorithm which does not use an exhaustive search through a structure (usually written as unbounded WHILE...DO, REPEAT...UNTIL, or $\mu$ operator); thus, it becomes possible to count working time of the algorithm. Although the upper complexity bounds for a PR-algorithm may be awfully large, this is a principal improvement compared with the general computability. As stressed in \cite{bd19}, PR-presentability of a structure may often be improved even to PTIME-presentability. 

Thus,  PR-presentability of structures (and PR-computability in general) seems important for the following reasons: it is in some respect close to feasible computability, is technically  easier than, say PTIME-computability, and has much better closure properties.
In this paper we investigate PR-versions of some  results  in \cite{ss17a,ss18}. The PR versions  have their own flavour and complement  the results in \cite{ss17a,ss18}. 

In particular, we find a PR-version (for Ahchimedean case) of the  Ershov-Madison theorem on the computable real closure \cite{er68,ma70}, relate  PR ordered fields of reals to the field of PR reals, propose (apparently, new) notions of PR-computability in analysis and apply them to obtain new results on computability of PDE-solutions. Our notion of PR Archimedean field of reals uses the idea of PR Skolem functions which was earlier used  by R.L. Goodstein in \cite{good61} in his development of a version of constructive analysis.\footnote{We are grateful to Vasco Brattka for the hint to Goodstein's monography.}   Our approach to PR computability in analysis sketched below is also related to the approach by W. Gomaa (see \cite{gom11}, comments after Remark 1) to PR computability on the reals.\footnote{We are grateful to an anonymous reviewer of \cite{ss21} for the hint to the survey by W. Gomaa.}

More specifically, we identify a class of  PRAS-fields (PR Archimedean fields of reals with PR-splitting) such that the above-mentioned problems over any such field are PR-computable.
These results complement and contrast the results in \cite{ss17a,ss18}, as well as some results in \cite{as18}. E.g., the class of PRAS-fields  is shown to be richer than the class of PTIME-presentable fields but the union of this class is a proper subset of the set of PR reals (in contrast with the corresponding fact in \cite{ss17a}). In the applications sections, we show how this applies to spectral problems and (PR) solving of symmetric hyperbolic systems of PDEs. For the latter example the solution is computed numerically (via difference schemes) while the algebraic part is performed symbolically; this part also requires introducing definitions of PR real functions and operators.

In programming terms, we identify an important class of number fields and related algorithmic problems of algebra and analysis which may be programmed without using the above-mentioned unbounded cycle operators.

After some preliminaries on computable and PR-structures in the next section and on PR rings and fields in Section \ref{prings}, we define in Section \ref{madis} PR-Archimedean fields and prove the PR-version of Ershov-Madison's theorem for such fields. In Section \ref{roots} we introduce PR-splitting property and give a sufficient condition for uniqueness of the PR real closure and for the existence of PR root-finding algorithms in the PR real and algebraic closures of PR-fields with this property.  In Section \ref{prreals} we examine  PR-versions of results in \cite{ss17,ss17a} on the relations of computable ordered fields of reals to the field of computable reals. In Section \ref{prtr} we show that many transcendental reals may be included in a PRAS-field, in contrast with the PTIME-presentable ordered fields of reals for which no such example is known. In Section \ref{algebra} we apply the obtained results to show the existence of PR algorithms for finding  normal forms of matrices over any PRAS-field frequently used in applications. In Section \ref{anal} we introduce basic notions of a PR version of computable analysis which seems closely related to Goodstein's version of constructive analysis. In Section \ref{prpde} we apply the obtained results to proving PR-computability of solutions operators for symmetric hyperbolic systems of PDE, complementing the results in \cite{ss17a,ss18}.  We conclude in Section \ref{open} with a short informal discussion.

\section{PR structures}\label{prel}

Here we recall some basic definitions and observations about PR functions and structures. We start with recalling  basic notions related to computable structures (see e.g. \cite{ma61,eg99,st99} for additional details).

A {\em numbering} is any function with domain $\nat$. Instead of $\nu(n)$ we may write $\nu_n$ whenever convenient. A {\em numbering of a set $B$} is a surjection $\beta$
from $\mathbb{N}$ onto $B$; sometimes we write $\beta_n$ or $\beta n$ instead of the ``canonical'' $\beta(n)$. For numberings $\beta$ and $\gamma$,
$\beta$ is {\em reducible} to $\gamma$ (in symbols
$\beta\leq\gamma$) iff $\beta=\gamma\circ f$ for some computable
function $f$ on $\mathbb{N}$, and $\beta$ is {\em equivalent} to
$\gamma$ (in symbols $\beta\equiv\gamma$) iff $\beta\leq\gamma$ and
$\gamma\leq\beta$.  These notions (due to A.N. Kolmogorov)
enable to transfer the computability theory over $\mathbb{N}$ to
computability theory over  many other countable structures.

\begin{Definition}\label{funpred}
Let $\nu:\nat\to B$  be a numbering. A relation $P\subseteq B^n$ on $B$ is $\nu$-computable if the relation $P(\nu(k_1),\ldots,\nu(k_n))$ on $\nat$ is computable. A function $f:B^n\to B$  is $\nu$-computable if  $f(\nu(k_1),\ldots,\nu(k_n))=\nu g(k_1,\ldots,k_n)$ for some  computable function $g:\nat^n\to \nat$; this notion is naturally extended to partial functions $f:\sq B^n\to B$. More generally, given another numbering $\mu:\nat\to C$, a function $f:B^n\to C$  is $(\nu,\mu)$-computable if  $f(\nu(k_1),\ldots,\nu(k_n))=\mu g(k_1,\ldots,k_n)$ for some  computable function $g:\nat^n\to \nat$.
\end{Definition}

\begin{Definition}\label{cons}
A structure $\mathbb{B}=(B;\sigma)$  of a finite signature $\sigma$
is called constructivizable iff there is a numbering $\beta$ of $B$
such that all signature predicates and functions, and also the
equality predicate,  are $\beta$-computable. Such a numbering
$\beta$ is called a constructivization of $\mathbb{B}$, and the pair
$(\mathbb{B},\beta)$ is called a constructive structure.
\end{Definition}

Note that all information about a constructive structure $(\mathbb{B},\beta)$ is contained in the numbering $\beta$ (and computable functions representing the equality and signature functions and relations in this numbering) because $B=\rng(\beta)$. We sometimes use this to shorten notation in Sections \ref{roots} and \ref{prreals}.

The notion of a constructivizable  structure is equivalent to the
notion of a computably presentable structure popular in the western
literature. Obviously, $(\mathbb{B},\beta)$ is  a constructive
structure iff, given  $n_1,\ldots,n_k\in\mathbb{N}$ and a quantifier-free $\sigma$-formula
$\phi(v_1,\ldots,v_k)$ with free variables among $v_1,\ldots,v_k$, one can compute the
value $\phi^\mathbb{B}(\beta(n_1),\ldots,\beta(n_k))$ of
$\phi$ in $\mathbb{B}$.

\begin{Definition}\label{scons}
A structure $\mathbb{B}=(B;\sigma)$ of a finite signature $\sigma$
is called strongly constructivizable iff there is a constructivization
$\beta$ of $B$ such that, given $n_1,\ldots,n_k\in\mathbb{N}$ and a first-order $\sigma$-formula
$\phi(v_1,\ldots,v_k)$, one can compute the
value $\phi^\mathbb{B}(\beta(n_1),\ldots,\beta(n_k))$. Such  $\beta$ is
called a strong constructivization of $\mathbb{B}$, and the pair
$(\mathbb{B},\beta)$ is called a strongly constructive structure.
\end{Definition}

Clearly, any strongly  constructivizible structure is
constructivizible and has a decidable first-order theory. The notion of a strongly constructive structure is equivalent to the notion of a decidable structure  popular in the
western literature.

PR-versions of the notions defined above are obtained by  changing ``computable'' to ``PR'' in the definitions above.   In particular, for numberings $\beta$ and $\gamma$,
$\beta$ is {\em PR-reducible} to $\gamma$ (in symbols
$\beta\leq_{PR}\gamma$) iff $\beta=\gamma\circ f$ for some PR
function $f$ on $\mathbb{N}$, and $\beta$ is {\em PR-equivalent} to
$\gamma$ (in symbols $\beta\equiv_{PR}\gamma$) iff $\beta\leq_{PR}\gamma$ and
$\gamma\leq_{PR}\beta$. For  $\nu:\nat\to B$, a relation $P\subseteq B^n$ on $B$ is $\nu$-PR if the relation $P(\nu(k_1),\ldots,\nu(k_n))$ on $\nat$ is PR. A function $f:B^n\to B$  is $\nu$-PR if  $f(\nu(k_1),\ldots,\nu(k_n))=\nu g(k_1,\ldots,k_n)$ for some  PR function $g:\nat^n\to \nat$. A structure $\mathbb{B}=(B;\sigma)$  
is {\em PR-constructivizable} iff there is a numbering $\beta$ of $B$
such that all signature predicates and functions, and also the
equality predicate,  are $\beta$-PR. Such 
$\beta$ is called a {\em PR-constructivization} of $\mathbb{B}$, and the pair
$(\mathbb{B},\beta)$ is called a {\em PR structure}.

Basic information on PR functions is supposed to be known; it may be found e.g. in \cite{ma64}. The PR functions are generated from the distinguished functions $o=\lambda n.0$, $s=\lambda n.n+1$, and $I^n_i=\lambda x_1,\ldots,x_n.x_i$ by repeated applications of the operators of superposition $S$ and primitive recursion $R$. Thus, any PR function is represented by a ``correct'' term in the partial algebra of functions over $\nat$ (``correct'' terms are those satisfying some obvious constraints on the arities of the involved functions which guarantee that the resulted function on $\nat$ is total). Intuitively, any total function defined by an explicit definition using (not too complicated) recursion is PR; the unbounded $\mu$-operator is of course forbidden but the bounded one is possible.

The PR-version of Definition \ref{cons} was introduced in \cite{ma61}. After a long break, PR-structures appeared as an intermediate stage of proving PTIME-presentability of a structure (see e.g. \cite{gr90,cr91}). After another break, PR-structures appeared in the literature as a promising candidate for capturing the so called online structures which give a practically relevant alternative to computable structures (see the discussion and additional references in \cite{bd19}).
The following notion from  \cite{bd19} is important for this approach: a structure $\mathbb{B}$ is {\em fully PR-presentable (FPR-presentable)} if it is isomorphic to a PR structure with universe $\nat$. 
Let us characterise the $FPR$-structures in our (i.e., Mal'cev's) terminology. We call a numbering $\nu$ {\em PR-infinite} if there is a PR function $f$ such that $\nu(f(i))\not=\nu(f(j))$ whenever $i\not=j$. 

\begin{Proposition}\label{fpr}
An infinite structure $\mathbb{B}$ is FPR-presentable iff it has a PR-constructi\-vi\-zation $\beta$ which is PR-infinite.
\end{Proposition}

{\em Proof.} We consider the less obvious direction. Let $\beta$ be a PR-constructivization which is PR-infinite via $f$. Define a function $g:\nat\to\nat$ as follows: $g(0)=0$ and $g(n+1)=\mu x.\forall i\leq n(\beta(x)\not=\beta(g(i)))$. Since $B$ is infinite, $g$ is total and injective. Let $h(n)=max\{f(0),\ldots,f(n)\}$. Then $g(n+1)=\mu x\leq h(n).\forall i\leq n(\beta(x)\not=\beta(g(i)))$ and  $h$ is PR, hence $g$ is also PR. The numbering $\gamma=\beta\circ g$ is  PR-reducible to $\beta$ and injective. Conversely, $\beta\leq_{PR}\gamma$ via the PR function $u(n)=\mu x\leq n.\beta(n)=\beta(x)$. Thus, $\gamma$ is a bijective numbering of $B$ PR-equivalent to $\beta$, so  it is a bijective PR-constructivization of $\mathbb{B}$. Copying  interpretations of signature symbols from $\mathbb{B}$ to $\nat$ via $\gamma^{-1}$ we obtain a PR-copy of $\mathbb{B}$ with universe $\nat$.  
 \qed

Note that any PR-constructivization $\beta$ of an associative commutative  ring with 1 of characteristic $0$ is PR infinite (via any PR function $f$ such that $\beta(f(i))$ coincide with the element $1+\cdots+1\in B$ ($i+1$ summands)). Since in the subsequent sections we consider mostly rings and fields of characteristic $0$,  most of the PR-constructivizable structures there are also FPR-presentable. As observed in \cite{km19}, in fact any PR-constructivizable infinite field is FPR-presentable. 

Importantly, all usual encodings and decodings of constructive objects (like pairs, triples, finite strings, terms, formulas and so on) used in computability theory and its applications may be done using PR functions \cite{ma64}. For instance, there is a PR bijection $\langle n_1,n_2\rangle$ between $\nat\times\nat$ and $\nat$, with  PR decoding functions. With some abuse, we use similar notation $\langle n_1,n_2,n_3\rangle$ to encode triples and finite strings of natural numbers. 

We give some examples of PR structures. The structure $(\nat;\leq,+,\cdot)$ is clearly PR. The ordered ring $\intt$ of integers is PR-constructivizable via the bijective numbering $\zeta(2n)=n$ and $\zeta(2n+1)=-n-1$. The ordered field $\rat$ of rationals is PR-constructivizable via the numbering $\varkappa(\langle n_1,n_2,n_3\rangle)=(n_1-n_2)(n_3+1)^{-1}$. This numbering may be in the obvious way improved to a  bijective PR-constructivization such that from a given $n$ one can primitive recursively compute a fraction $a/b=\varkappa()$, $a\in\intt,b\in\nat$, and vice versa; below we always assume that   $\varkappa$ is a bijection with these properties..

Let $\{t_n\}$ be the G\"odel numbering of all variable-free terms of  signature $\sigma\cup\{c_0,c_1,\ldots\}$ where $c_0,c_1,\ldots$ is an infinite sequence of new constant symbols. Let $\mathbb{B}$ be a $\sigma$-structure. We associate with any numbering $\nu:\nat\to B$ the numbering $\tilde{\nu}:\nat\to B$ as follows: $\tilde{\nu}(n)$ is the value of $t_n$ in $\mathbb{B}$ when the constant $c_i$ is interpreted as $\nu(i)$. The following fact is a straightforward PR-analogue of  a particular case of a theorem in \cite{s82}.

\begin{Proposition}\label{gen}
\begin{enumerate}\itemsep-1mm
\item For all $\mu,\nu:\nat\to B$ we have: $\nu\leq_{PR}\tilde{\nu}$, $\mu\leq_{PR}\nu$ implies $\tilde{\mu}\leq_{PR}\tilde{\nu}$, and $\tilde{\tilde{\nu}}\leq_{PR}\tilde{\nu}$; any $\sigma$-function is $\tilde{\nu}$-PR; if  $\nu\leq_{PR}\mu$ and any $\sigma$-function is $\mu$-PR then $\tilde{\nu}\leq_{PR}\mu$.
\item If $(\mathbb{B},\beta)$ is a PR $\sigma$-structure and $\nu\leq_{PR}\beta$ then $\tilde{\nu}\leq_{PR}\beta$ and $(\rng(\tilde{\nu}),\tilde{\nu})$ is a PR $\sigma$-substructure of $(\mathbb{B},\beta)$ generated by $\rng(\nu)$; in particular, $\tilde{\beta}\equiv_{PR}\beta$.
\end{enumerate}
\end{Proposition}

We conclude this section with recalling a nice characterization of unary PR functions due to R. Robinson (see e.g. Section 3.5 in \cite{ma64}). Consider the structure $(\calN;+,\circ,J,\mathbf{s},\mathbf{q})$  where $\calN=\nat^\nat$ is the set of unary functions on $\nat$, $+$ and $\circ$ are binary operations on $\calN$ defined by $(p+q)(n)=p(n)+q(n)$ and $(p\circ q)(n)=p(q(n))$, $J$ is a unary operation on $\calN$ defined by $J(p)(n)=p^n(0)$ where $p^0=id_\nat$ and $p^{n+1}=p\circ p^n$, $\mathbf{s}$ and $\mathbf{q}$ are distinguished elements defined by $\mathbf{s}(n)=n+1$ and $\mathbf{q}(n)=n-\lfloor\sqrt{n}\rfloor^2$ where $\lfloor x\rfloor$ is the integer part of  $x\in\IR$, i.e., the unique integer $m$ with $m\leq x<m+1$. This structure is known as the Robinson algebra.

Let $T_0$ be the set of variable-free terms $t$ of signature $\tau=\{+,\circ,J,s,q\}$. The value $\bft$ of $t$ in the Robinson algebra is an element of $\calN$. By  Robinson's theorem, the set $\{\bft\mid t\in T_0\}$ of values $\bft$ of such terms  coincides with the set of unary PR functions. Using a standard G\"odel's encoding  $N:T_0\to\nat$ by natural numbers (e.g., we can set $N(\mathbf{s})=1$, $N(\mathbf{q})=3$, $N(t+u)=2\cdot3^{N(t)}\cdot5^{N(u)}$,$N(t\circ u)=4\cdot3^{N(t)}\cdot5^{N(u)}$, and $N(J(t))=8\cdot3^{N(t)}$), one can construct a computable  numbering $\psi$ of the unary PR functions such that the operations $+,\circ,J$ are $\psi$-PR (see e.g. Section 5.2 of \cite{ma64} for the details). Recall that computability of $\psi$ means that the binary function $\psi(n)(x)$ is computable; this binary function is  not PR. Moreover there is a sequence $\{t_k\}$ in $T_0$ such that $\bft_k=\psi(k)$ and the function $k\mapsto N(t_k)$ is PR. 

More generally, for any $n\geq0$, let $T_n$ be the set of    terms $t=t(v_1,\ldots,v_n)$ of signature $\tau$ with variables among a fixed list $v_1,\ldots,v_n$ of pairwise distinct variables. Any $t\in T_n$ determines the $n$-ary operator $\bft$ on $\calN$ by setting $\bft(g_1,\ldots,g_n)$ to be the value of $t$ for $v_i=g_i$. The functions of the form $\bft(g_1,\ldots,g_n)$ coincide with the functions PR in $g_1,\ldots,g_n$ (the last notion is obtained from the ``correct'' terms mentioned in the Introduction by adding $g_1,\ldots,g_n$ to the list of basic functions). As above, this yield a G\"odel numbering $\psi^n$ of {\em $n$-ary PR operators on $\calN$}. For $n=0$ this numbering coincides with the numbering $\psi$ of PR functions in the preceding paragraph. For $n=1$ we obtain a numbering of unary PR-operators on $\calN$ which is PR-reducible to the standard numbering of Turing functionals. We use the introduced  PR-operators on $\calN$ in Section \ref{anal} to define PR functions on the reals.

\section{PR rings and fields}\label{prings}

 The literature on computable rings and fields is very rich but, surprisingly, the PR analogue of this theory does not seem to be considered seriously so far. In this section we collect a few notions and facts on PR rings and fields which are used in the sequel and which are analogues of their computable versions. We assume the reader to be familiar with  basic notions and facts about rings and fields (see e.g. \cite{wa67}). The word ``ring'' in this paper usually means ``associative commutative ring with 1'', the only exception are  rings of matrices which are in general not commutative.
The investigation of PR rings and fields in parallel to the well known corresponding computable versions \cite{eg99,st99} is mostly straightforward but from time to time some complications do appear. 
 We consider fields and ordered  fields  in  signatures $\{+,\cdot,-,^{-1},0,1\}$ and $\{+,\cdot,-,^{-1}\leq,0,1\}$ resp.; for (ordered) rings the symbol $^{-1}$ is of course removed.

Recall that an {\em integral domain} is a ring  with no zero-divisors satisfying $1\not=0$. With any integral domain $\mathbb{B}$ we  associate its quotient field $\widetilde{\mathbb{B}}$. The elements of $\widetilde{B}$ are the equivalence classes $[a,b]$ of pairs $(a,b)$ where $a,b\in B,b\not=0$ and the equivalence on pairs is defined by: $(a,b)\sim(a',b')$ iff $ab'=ba'$. With any numbering $\beta$ of $B$ we associate the numbering $\tilde{\beta}$ of $\tilde{B}$ as follows: $\tilde{\beta}(\langle n_1,n_2\rangle)=[\beta(n_1),\beta(n_2)]$ if $\beta(n_2)\not=0$, and $\tilde{\beta}(\langle n_1,n_2\rangle)=0$ otherwise. The next proposition is straightforward.

\begin{Proposition}\label{qfield}
\begin{enumerate}\itemsep-1mm
\item If $\beta$ is a PR-constructivization of an integral domain $\mathbb{B}$ then $\tilde{\beta}$ is a PR-constructivization of the quotient field $\widetilde{\mathbb{B}}$.
\item If $\beta$ is a PR-constructivization of an ordered ring $\mathbb{B}$ (which is automatically an integral domain) then $\tilde{\beta}$ is a PR-constructivization of the quotient ordered field $\widetilde{\mathbb{B}}$ (with the unique ordering extending the given ordering of $\mathbb{B}$).
\end{enumerate}
 \end{Proposition}

With any numbering $\beta$ of a (countable) ring $\mathbb{B}$ we associate the numbering $\beta^*$ (the precise notation should refer to the variable $x$, say we could use $\beta^{[x]}$) of the ring $\mathbb{B}[x]$ of polynomials over $\mathbb{B}$ with variable $x$ as follows: $\beta^*(\langle i_0,\ldots,i_n\rangle)=\beta(i_0)x^0+\cdots+\beta(i_n)x^n$ where $\langle i_0,\ldots,i_n\rangle$ is a PR coding of the set $\nat^+=\bigcup_n\nat^{n+1}$ of finite non-empty strings of natural numbers. 
Iterating this construction, we obtain for each $n$ the numbering $\beta^{[n]}$ of $\mathbb{B}[x_0,\ldots,x_n]$ (identifying  $\mathbb{B}[x_0,\ldots,x_{n+1}]$ with $\mathbb{B}[x_0,\ldots,x_n][x_{n+1}]$) as follows: $\beta^{[0]}=\beta^*$, $\beta^{[n+1]}=(\beta^{[n]})^*$. Then the expression $\beta^{[\omega]}(\langle n,m\rangle)=\beta^{[n]}(m)$ defines a numbering of the set $\mathbb{B}[x_0,x_1,\ldots]$ of polynomials with any number of variables. The next proposition is straightforward.

\begin{Proposition}\label{polynom}
\begin{enumerate}\itemsep-1mm
 \item Let $(\mathbb{A},\alpha)$ and $(\mathbb{B},\beta)$ be PR rings, $\mathbb{A}$  a subring of $\mathbb{B}$, and $\alpha\leq_{PR}\beta$. Then $\alpha^{[n]}\leq_{PR}\beta^{[n]}$ for all $n$.
 \item If $\beta$ is a PR-constructivization of $\mathbb{B}$ then $\beta^{[0]},\beta^{[1]}\ldots,\beta^{[\omega]}$ are PR-constructivizations of rings $\mathbb{B}[x_0],\mathbb{B}[x_0,x_1],\ldots,\mathbb{B}[x_0,x_1,\ldots]$ respectively.
\end{enumerate}
 \end{Proposition}

For any $n$, let $ev_n:\mathbb{B}[x_0,\ldots,x_n]\times \mathbb{B}^n\to\mathbb{B}$ be the evaluation function defined as follows: $ev_n(p,b_0,\ldots,b_n)$ is the value $p(b_0,\ldots,b_n)$ of $p$ at $(b_0,\ldots,b_n)\in\mathbb{B}^n$. Let $ev_\omega:\mathbb{B}[x_0,x_1,\ldots]\times \mathbb{B}^+\to\mathbb{B}$ be the evaluation function which returns the value of a polynomial if the arities match and returns $0$ otherwise.  The next proposition is straightforward.

\begin{Proposition}\label{eval}
If  $(\mathbb{B},\beta)$ is a PR ring then the evaluation functions are PR  in the corresponding PR-constructivizations of the polynomial rings above.
 \end{Proposition}

Polynomials are closely related to terms in the signature of rings. We formulate a straightforward fact about such a relation to terms in signature $\sigma=\{+,\cdot,c_0,c_1,\ldots\}$. The direction from polynomials to terms is checked by induction on the number of monomials, the opposite direction --- by induction on (the rank of) terms. Interestingly, the PTIME-version of this fact fails in both directions.

\begin{Proposition}\label{polterm}
Let $(\mathbb{B},\beta)$ be a PR ring. Given $p\in\mathbb{B}[x_0,\ldots,x_n]$, one can primitive recursively find a $\sigma$-term $t(x_0,\ldots,x_n)$ such that $p(b_0,\ldots,b_n)=t^\mathbb{B}(b_0,\ldots,b_n)$ for all $b_0,\ldots,b_n\in B$, where $c_n$ is interpreted as $\beta(n)$. The same holds for the opposite direction $t\mapsto p$.
 \end{Proposition}

Below we consider  rings of polynomials only over integer domains and over fields. As is well known, the ring $\mathbb{B}[x]$  has interesting arithmetic resembling in many respects the integer arithmetic. We mention (not systematically) some terminology and algorithms which have natural PR analogues. If $ p = a_{n}x^{n} + \ldots + a_{1}x + a_{0}\in\mathbb{B}[x] $,
where $ a_{n} \neq 0 $, then $ n $ is called the {\it degree of $p$},  denoted as
$ \deg (p) $. The element $ a_{n} $ is the {\it leading coefficient} of $ p  $.
If $ a_{n} = 1 $ or ($ n = 0 $ and $ a_{n} = 0 $), then the polynomial is called {\it normalized}.
We say that $ p_{2}  $ {\it divides} $ p_{1}  $ in
$ \mathbb{B}[x] $,
$ p_{2} (x) \dv p_{1} (x) $, if there is $ q  \in\mathbb{B} [x] $ with
$ q  p_{2}  = p_{1}  $. A polynomial $ q\in\mathbb{B} $
is a {\it greatest common divisor of}
$ p_{1},p_{2} $, if $ q \dv p_{i} $ for $ i = 1,2 $ and the conditions
$ r  \dv p_{i}  $ for $ i = 1,2 $ imply $ r  \dv q $. A greatest common divisor always exists and
is unique up to multiplication by a non-zero constant in $\mathbb{B} $.
We denote by
$ \gcd (p_{1} , p_{2}) $
the normalized greatest common divisor.

A polynomial $ p \in \mathbb{B} [x] $ over a field is  {\it irreducible} in $ \mathbb{B} [x] $
if $ \deg (p) \gs 1 $ and the equality $ p  = p_{1}  p_{2}  $ implies that
$ \deg (p_{1}) = 0 $ or $ \deg (p_{2}) = 0 $. If $ p  $ is an irreducible polynomial and
$ p  \dv p_{1}  p_{2}  $, then $ p  \dv p_{1}  $ or $ p  \dv p_{2} $. Every non-zero polynomial
$ p  \in \mathbb{B} [x] $ is representable in the form
$ p  = a p_{1}  p_{2}  \ldots p_{k}  $, where $ a \in \mathbb{B} $ and $ p_{i}  $
are normalized irreducible polynomials, and such a representation is unique up to factor permutation; it is called the {\it canonical decomposition} of $p$.

The algorithm of polynomial division of $p_1$ by $p_2$  (where $p_1,p_2\in\mathbb{B}[x]$ with $\degr(p_1)\geq\degr(p_2)>0$ are given polynomials over a field)   returns the quotient $q=\quot(p_1,p_2)$ and the remainder $r=\rest(p_1,p_2)$ polynomials such that $p_1=qp_2+r$ and $\deg(r)<\deg(p_2)$. The Euclidean algorithm, applied to  such polynomials $p_1,p_2$, produces an iterated sequence of quotients and remainders  and returns the  greatest common divisor $\gcd(p_1,p_2)$ of $p_1,p_2$ as the last non-zero remainder. 

The {\it derivative} $ p'  $ of a polynomial
$ p  = a_{n}x^{n} + a_{n-1}x^{n-1} + \ldots + a_{2}x^{2} + a_{1}x + a_{0} $
is defined as
$ na_{n}x^{n-1} + (n-1)a_{n-1}x^{n-2} + \ldots + 2a_{2}x + a_{1} $.
If
$ p  = a p_{1}^{e_{1}} p_{2}^{e_{2}} \ldots p_{k}^{e_{k}} $,
where $ a \in \mathbb{B} $, $ p_{i}  $ are distinct normalized irreducible polynomials,
$ e_{i} \gs 1 $, and $ p_{i} (x) \neq p_{j} (x) $ for $ i \neq j $, then
$ \gcd (p, p') = p_{1}^{e_{1}-1} p_{2}^{e_{2}-1} \ldots p_{k}^{e_{k}-1} $, hence $ p  / \gcd (p (x), p') = ap_{1} \ldots p_{k} $.
A polynomial $ p  $ is {\it square free} if in the factorization above $ e_{i} = 1 $
for $ i \ls k $. For polynomials over a field $\mathbb{B} $ of characteristic 0 this is equivalent to
$ \gcd (p, p') = 1 $. In every field that extends $\mathbb{B} $, the roots of
$ p  $ and $ p  / \gcd (p, p') $ coincide and the latter polynomial does not have multiple roots.
 The next proposition is straightforward.

\begin{Proposition}\label{arith}
If $(\mathbb{B},\beta)$ is a PR field  then $(\mathbb{B}^*,\beta^*)$ is a PR integral domain and the functions $\deg,\quot,\rest,\gcd,'$ and the relations $\dv$, ``$p_1,p_2$ are relatively prime'',  ``$p$ is square-free'' are $\beta^*$-PR.
 \end{Proposition}

For any field $\mathbb{B}$, $\mathbb{B}[x_0,\ldots,x_n]$ is an integral domain whose quotient field $\mathbb{B}(x_0,\ldots,x_n)$ is known as the field of rational functions over $\mathbb{B}$; let $g:\mathbb{B}[x_0,\ldots,x_n]^2\to\mathbb{B}(x_0,\ldots,x_n)$ be the surjection essentially defined before Proposition \ref{qfield}. For any normalized irreducible $p\in\mathbb{B}[x]$, let $\mathbb{B}/_{(p)}$ be the quotient field of $\mathbb{B}[x]$ by the maximal ideal $(p)=\{q\in\mathbb{B}[x]: p\dv q\}$ of $\mathbb{B}[x]$, and let $g_p:\mathbb{B}[x]\to\mathbb{B}/_{(p)}$ be the canonical surjection.  The next proposition follows easily from the previous one and Proposition \ref{qfield}.

\begin{Proposition}\label{rfunc}
If $(\mathbb{B},\beta)$ is a PR field  then so are also $(\mathbb{B}(x_0,\ldots,x_n),g\circ(\beta^{[n]})^2)$ and $(\mathbb{B}/_{(p)},g_p\circ\beta^*)$.
 \end{Proposition}

In the study of computable fields, the property of computable splitting  is important. The PR-version of this notion is as follows: a PR field $(\mathbb{B},\beta)$ {\em has PR splitting} if, given $p\in\mathbb{B}[x]$, one can primitive recursively find a canonical decomposition of $p$. Note that this implies that the property of being a reducible polynomial is $\beta^*$-PR. We do not currently know whether the opposite implication holds (although its computable version obviously holds).
The following assertion collects straightforward PR-analogues of the corresponding well known facts about computable splitting \cite{eg99,st99}.

\begin{Proposition}\label{prsplit}
\begin{enumerate}\itemsep-1mm
\item The PR field $(\rat,\varkappa)$ has PR splitting.
\item If $(\mathbb{B},\beta)$ is a PR field which has PR splitting then so are  the fields $(\mathbb{B}(x_0,\ldots,x_n),g\circ\beta^{[n]})$ and $(\mathbb{B}/_{(p)},g_p\circ\beta^*)$ from the previous proposition.
\end{enumerate}
 \end{Proposition}
 
Let  $\mathbb{A}\subseteq\mathbb{B}$ be a field extension, $b_0,\ldots\in B\setminus A$, and let $\mathbb{A}(b_0,\ldots)$ be the subfield of $\mathbb{B}$ generated by $A\cup\{b_0,\ldots\}$. If $b_0,\ldots$ are algebraically independent (hence transcendental over $\mathbb{A}$) then $\mathbb{A}(b_0,\ldots)$ is known to be isomorphic to the field $\mathbb{A}(x_0,\ldots)$ of rational functions (the standard isomorphism is induced by the polynomial evaluation function). If $b_0$  is algebraic over $\mathbb{A}$ then $\mathbb{A}(b_0)$ is known to be isomorphic to the field $\mathbb{A}[x]/_{(p_{b_0})}$ where $p_{b_0}\in\mathbb{A}[x]$ is the (unique) normalized polynomial of minimal degree with $p_{b_0}(b_0)=0$. 

If $\mathbb{A}$ has characteristic $0$ and $\mathbb{B}$ is an algebraic closure of $\mathbb{A}$ then for any  $b_0,\ldots,b_n\in B$  there exists  $b\in\mathbb{B}$ (called a primitive element for $b_0,\ldots,b_n$) such that $\mathbb{A}(b)=\mathbb{A}(b_0,\ldots,b_n)$   (this follows from the well known primitive element theorem, see e.g. Section 46 of \cite{wa67}).  Theorem 11 in \cite{lo82} provides a constructive version of this theorem. Item 3 of the following proposition is a  PR-version of this constructive version with essentially the same proof. Items 1 and 2 are straightforward PR versions of the facts mentioned in the previous paragraph.

\begin{Proposition}\label{prmitive}
Let $(\mathbb{A},\alpha)$ and $(\mathbb{B},\beta)$ be PR fields such that $\mathbb{A}\subseteq\mathbb{B}$, $\alpha\leq_{PR}\beta$, $(\mathbb{A},\alpha)$ has PR splitting, and let $b_0,\ldots\in B\setminus A$ be a finite or infinite sequence which is PR reducible to $\beta$ provided it is infinite. 
\begin{enumerate}\itemsep-1mm
\item If $b_0,\ldots$ are algebraically independent over $\mathbb{A}$ then the standard isomorphism between $\mathbb{A}(b_0\ldots)$ and  $\mathbb{A}(x_0,\ldots)$, as well as  its inverse, are represented by PR functions on the names.
\item If $b_0$  is algebraic over $\mathbb{A}$ then the standard isomorphism between $\mathbb{A}(b_0)$ is  $\mathbb{A}[x]/_{(p_{b_0})}$ and its inverse are represented by PR functions on the names.
\item Let $\mathbb{A}$ have characteristic $0$ and let $\mathbb{B}$ be an algebraic closure of $\mathbb{A}$. Given $p_0,\ldots,p_n\in\mathbb{A}[x]$ with $p_i(b_i)=0$ for  $i\leq n$, one can primitive recursively find a primitive element $b$ for $b_0,\ldots,b_n$ and polynomials $p,q_0,\ldots,q_n\in\mathbb{A}[x]$ such that $p(b)=0$ and $q_i(b)=b_i$ for  $i\leq n$.
\end{enumerate}
 \end{Proposition}

Recall that a polynomial in $\mathbb{B}[x_1,\ldots,x_n]$ is {\em symmetric} if it does not change under any permutation of the variables. The symmetric polynomials
\begin{eqnarray*}
s_1=x_1+\cdots +x_n,\;s_2=x_1x_2+x_1x_3+\cdots +x_{n-1}x_n,\\
s_3=x_1x_2x_3+x_1x_2x_4+\cdots +x_{n-2}x_{n-1}x_n,\;\ldots,\; s_n=x_1x_2\cdots x_n
\end{eqnarray*}
 are called {\em elementary}. It is well known that any  symmetric polynomial  $p\in\mathbb{B}[x_1,\ldots,x_n]$ may be obtained by substituting the expressions for $s_1,\ldots,s_n$ in a suitable polynomial $\varphi\in\mathbb{B}[s_1,\ldots,s_n]$. The standard proof of this fact (see e.g. Section 33 in \cite{wa67}) is constructive, and in fact the algorithm in that proof is PR. This yields the following.

\begin{Proposition}\label{symmetric}
Let $(\mathbb{B},\beta)$ be a PR ring. Given a symmetric   $p\in\mathbb{B}[x_1,\ldots,x_n]$, one can primitive recursively compute  a  polynomial $\varphi\in\mathbb{B}[s_1,\ldots,s_n]$ such that  $p=\varphi(s_1,\ldots,s_n)$.
 \end{Proposition}

Let $M_n(\mathbb{B})$ be the ring of $n\times n$-matrices over a ring $\mathbb{B}$. We use standard terminology and notation from linear algebra. In particular, ${\rm det}(A)$ is the determinant of $A=(a_{ij})\in M_n(\mathbb{B})$, ${\rm diag}(a_1,\ldots,a_n)$ is the diagonal matrix with  diagonal elements $a_1,\ldots,a_n\in R$, so in particular $I=I_n={\rm diag}(1,\ldots,1)$ is the unit matrix. The polynomial $ch_A={\rm det}(\lambda I-A)\in\mathbb{B}[\lambda]$ is called the {\em characteristic polynomial} of $A$. With any numbering $\beta$ of $\mathbb{B}$ we associate the numbering $\mu_n$ of  $M_n(\mathbb{B})$, by using the PR-bijection between $\nat^{n\times n}$ and $\nat$. Let $G$ be the set of {\em non-degenerate} matrices $A\in M_n(\mathbb{B})$, i.e. matrices with non-zero determinant. If $\mathbb{B}$ is a field then $G$ coincides with the set of matrices $A$ for which the matrix $A^{-1}$ exists. Also rectangular matrices are useful, in particular for solving general (i.e., with arbitrary number of variables and equations) linear systems of equations. The next proposition is straightforward.

\begin{Proposition}\label{matrix}
Let $(\mathbb{B};\beta)$ be a PR ring.
\begin{enumerate}\itemsep-1mm
\item For any $n>0$, $\mu_n$ is a PR-constructivization of the ring $M_n(\mathbb{B})$ uniformly in $n$.
\item The function ${\rm det}:M_n(\mathbb{B})\to\mathbb{B}$ is  $(\mu_n,\beta)$-PR, and the set $\mu_n^{-1}(G)$ is PR uniformly in $n$.
\item The function $ch:M_n(\mathbb{B})\to\mathbb{B}[\lambda]$ is  $(\mu_n,\beta^*)$-PR  uniformly in $n$.
\item If $\mathbb{B}$ is a field then  there  is a PR function $f:\nat\times\nat\to\nat$ such that $(\mu_n(i))^{-1}=\mu_n(f(i,n))$ for all $n$ and $i\in\mu_n^{-1}(G)$.
\item Given a general linear system of equations over a PR field, one can (using, say, the Gauss method) primitive recursively find  its general solution (i.e., a fundamental system of solutions).
\end{enumerate}
 \end{Proposition}
 
Let $\mathbb{B}$ be a ring and $p=a_0x^0+\cdots+a_mx^m$, $q=b_0x^0+\cdots+b_nx^n$ be polynomials over $\mathbb{B}$. Recall that  {\em Silvester matrix for $p,q$} is the matrix $S=S(p,q)\in M_{m+n}(\mathbb{B})$ whose first $n$ rows are  $$(a_m,\ldots,a_0,0,\ldots,0), (0,a_m,\ldots,a_0,0,\ldots,0), (0,0,a_m,\ldots,a_0,0,\ldots,0),\ldots$$ 
and whose last $m$ rows are  
 \[(b_n,\ldots,b_0,0,\ldots,0), (0,b_n,\ldots,b_0,0,\ldots,0), (0,0,b_n,\ldots,b_0,0,\ldots,0),\ldots\]  

\begin{Definition}\label{result}
For $m,n>0$, the resultant of $p,q$, denoted $\res(p,q)$ is defined as the determinant $\det(S)$ of the Silvester matrix for $p,q$. We also define $\res(p,q)=a_0^n,b_0^m,1$ respectively for $m=0<n$, $n=0<m$,  $m=0=n$.
\end{Definition}

Thus, $\res:\mathbb{B}[x]\times\mathbb{B}[x]\to\mathbb{B}$. Note that the resultant may also be considered as a fixed polynomial in $\intt[a_0,\ldots,a_m,b_0,\ldots,b_n]$. 
Clearly, if $\beta$ is a PR-constructivization of $\mathbb{B}$ then the resultant is a PR function in the corresponding numberings $\beta^*,\beta$. 

Till the end of this section we  assume that $\mathbb{B}$ is a field. Recall that  {\em discriminant $D(p)$ of $p=a_0x^0+\cdots+a_mx^m\in\mathbb{B}[x]$} (where $a_m\neq0$ and $m>1$) is defined by $D(p)=a_m^{2m-2}\prod_{i< j}(x_i-x_j)^2$ where $x_1,\ldots,x_m$ is a list of all roots (possibly with repetition) of $p$ in an algebraic closure $\overline{\mathbb{B}}$ of $\mathbb{B}$. In the next proposition we recall some properties of  resultant and discriminant from \cite{wa67,lo82}. Some other properties proved in \cite{lo82} will be recalled  later.

\begin{Proposition}\label{res}
\begin{enumerate}\itemsep-1mm
\item We have: $\res(p,q)=0$ iff either $a_m=b_n=0$ or $\gcd(p,q)$ has degree $>0$.
\item Let $p$ be a polynomial over $\mathbb{B}$ of degree $m>1$. Then $p$ has a multiple root in $\overline{\mathbb{B}}$ iff $D(p)=0$ iff $p$ is not square-free iff $\gcd(p,p')\neq1$.
\item Let $p$ be a polynomial over $\mathbb{B}$ of degree $m>1$. Then $res(p,p')=\pm a_mD(p)$,  hence $D(p)\in\mathbb{B}$.
\item Let $p=a_m\prod_i(x-x_i)$ and $q=b_n\prod_j(x-y_j)$ be  in $\mathbb{B}[x_1,\ldots,x_m,y_1,\ldots,y_n][x]$. Then $\res(p,q)=(-1)^{mn}b_n^m\prod_jp(y_j)=a_m^n\prod_iq(x_i)=a_m^nb_n^m\prod_{i,j}(x_i-y_j)$.
\item Let  $p,q$ be polynomials of positive degrees $m,n$ over $\mathbb{B}$, and  $x_1,\ldots,x_m$ and $y_1,\ldots,y_n$ be lists of all roots of $p$ and $q$ in $\overline{\mathbb{B}}$, respectively. Then the equalities of the previous item hold.
\end{enumerate}
 \end{Proposition}
 
As already mentioned,  proofs of results in this section follow easily from   known proofs.  PR analogues of many other algebraic facts may also be obtained this way. Below we prove some less straightforward results from PR algebra and analysis.

\section{PR real closure}\label{madis}

By a classical theorem of Artin and Schreier, for any ordered field $\mathbb{A}$
there exists an algebraic ordered extension $\widehat{\mathbb{A}}\supseteq\mathbb{A}$ which is real closed. Such an extension, called the {\em real closure of $\mathbb{A}$}, is unique in the sense that for any real closure $\mathbb{B}$ of $\mathbb{A}$ there is a unique isomorphism between $\mathbb{B}$ and $\widehat{\mathbb{A}}$ identical on $\mathbb{A}$.

Yu.L. Ershov \cite{er68} and independently E.W. Madison \cite{ma70} proved a computable version of the Artin-Schreier theorem: if $\mathbb{A}$ is constructivizable then so is also $\widehat{\mathbb{A}}$.
In this section we make search for a PR analogue of the Ershov-Madison theorem. Though we have not found a complete analogue, we describe one for the case of Archimedean ordered fields.
We recall some details of the proof of Ershov-Madison's theorem in \cite{ma70} which are used for our PR version  (the proof  in \cite{er68} is a particular case of a proof of a  more general model-theoretic result and it is not currently clear how to adjust it to the PR case). 

The author of \cite{ma70} defines, from any given constructivization $\alpha$ of an ordered field $\mathbb{A}$, a constructivization $\widehat{\alpha}$ of the real closure $\widehat{\mathbb{A}}$ as follows (we use slightly  modified notation). 
Let $P(i,k)$ mean that either $\alpha^*_i$ is the zero polynomial (i.e., all coefficients of $\alpha^*_i$ are zero) or $\alpha^*_i$  has at most $k$  roots in $\widehat{\mathbb{A}}$.
Then $\widehat{\alpha}(\langle i,k\rangle)=0$  if $P(i,k)$, otherwise $\widehat{\alpha}(\langle i,k\rangle)$  is the $(k+1)$-st (w.r.t. $<$)  root $b$ of $\alpha^*_i$ in $\widehat{\mathbb{A}}$ (i.e., $\alpha^*_i(b)=0$ and there are precisely $k$  roots of $\alpha^*_i$ in $\widehat{\mathbb{A}}$ strictly below $b$). 

It remains to check that $+,\cdot,\leq$ are $\widehat{\alpha}$-computable. As observed in \cite{ma70}, this follows  from the Tarski quantifier elimination  for real closed fields \cite{ta51}. For instance, in order to check $\widehat{\alpha}$-computability of $+$  (equivalently, computability of the relation $\widehat{\alpha}(m)+\widehat{\alpha}(n)=\widehat{\alpha}(p)$ on $\nat$) it suffices to note that, using the definition of $\widehat{\alpha}$ one can compute, given $m,n,p\in\nat$, a first-order formula $\phi(x_0,\ldots,x_l)$ of signature $\sigma=\{+,\cdot,-,^{-1},\leq,0,1,c_0,c_1,\ldots\}$ and a string  $(j_0,\ldots,j_l)$ of natural numbers such that $\widehat{\alpha}(m)+\widehat{\alpha}(n)=\widehat{\alpha}(p)$ iff  $\phi(\alpha(j_0),\ldots,\alpha(j_l))$ is true in $\widehat{\mathbb{A}}$. By the quantifier elimination, we can think that $\phi$ is quantifier-free. Since $\mathbb{A}$ is a substructure of $\widehat{\mathbb{A}}$, the values of $\phi(\alpha(j_0),\ldots,\alpha(j_l))$ in both structures coincide. Since $(\mathbb{A},\alpha)$ is constructive, the relation $\widehat{\alpha}(m)+\widehat{\alpha}(n)=\widehat{\alpha}(p)$ is computable.  

Unfortunately, this argument does not automatically yield the PR-version. 
If $(\mathbb{A},\alpha)$ is PR  then the Madison proof does yield that the graphs of signature functions are PR in  $\widehat{\alpha}$. This follows from the proof sketched above because Tarski's quantifier elimination is PR, i.e., from an arbitrary first-order formula $\phi$ as above one can primitive recursively find an equivalent quantifier-free formula (even better complexity bounds for quantifier elimination are known, see e.g. \cite{bp06} and references therein). So, at first glance the PR-version of the Ershov-Madison theorem automatically follows from Madison's proof sketched above. But primitive recursiveness of  a graph of a function  does not imply primitive recursiveness of the function itself, see e.g. Section 6.4 of \cite{ma64}. This is a major obstacle to proving the PR-version of the Ershov-Madison theorem in full generality. In fact, we currently do not have a proof for the general case. But we do have a reasonable version for a natural subclass of Archimedean fields.

Every Archimedean ordered field is isomorphic to a substructure of the ordered field $\IR$ of reals, so in the sequel we always assume that $\mathbb{A}\subseteq\IR$. In fact, we can also assume that $\widehat{\mathbb{A}}\subseteq\IR$ since $\widehat{\mathbb{A}}$ is isomorphic (over $\mathbb{A}$) to the ordered field of real roots of non-zero polynomials from $\mathbb{A}[x]$. So, from now on we always assume that $\mathbb{A}$ and $\widehat{\mathbb{A}}$ are ordered subfields of $\IR$, i.e. $\alpha,\widehat{\alpha}:\nat\to\IR$, and $\widehat{A}$ is the set of real roots of non-zero polynomials over $\mathbb{A}$.

Note that if $\alpha:\nat\to\IR$ is a constructivization of $\mathbb{A}$ then $(\mathbb{A},\alpha)$ is computably Archimedean in the sense that $\alpha(n)\leq f(n)$, for a suitable computable function $f$. Again, the PR-version of the last fact, probably, does not hold in general. In fact, our proof below  works only for {\em PR-Archimedean  fields} which we define as the  PR ordered subfields $(\mathbb{A},\alpha)$ of $\IR$ such that there is a PR function $f$ with $\forall(\alpha(n)\leq f(n))$. The main result of this section is the following.

\begin{Theorem}\label{main1}
If $(\mathbb{A},\alpha)$ is a PR-Archimedean  subfield of $\IR$ then so is also $(\widehat{\mathbb{A}},\widehat{\alpha})$.
\end{Theorem}

The theorem follows from facts 1 --- 7 proved below. Most of the facts just show that some standard algebraic functions are PR. The proofs of some facts are   versions of the corresponding proofs in \cite{lo82,as18,as19} which show that the ordered field $\ra$ of algebraic reals is PTIME-presentable.
 \medskip

{\bf Fact 1.} There is a PR function $f$ such that all real roots of any non-zero polynomial $\alpha^*_i\in\mathbb{A}[x]$ are in the interval $(-f(i),f(i))$. In particular, $\widehat{\alpha}(\langle i,k\rangle)<f(i)$ for all $i,k$ (so $(\widehat{\mathbb{A}},\widehat{\alpha})$ is PR-Archimedean provided that it is a PR ordered field).
 \medskip
 
{\em Proof.} By the notation in Section \ref{prings}, $\alpha^*_i=\alpha(i_0)x^0+\ldots+\alpha(i_n)x^n$. Since $\alpha^*_i$ is non-zero, we have $\alpha(i_m)\not=0$ where $m\leq n$ is the degree of $\alpha^*_i$. As is well known, all real roots of $\alpha^*_i$ are in $(-M_i;M_i)$ where $M_i=1+a|\alpha(i_m)^{-1}|\in A$ and $a=max\{|\alpha(i_j)|:j<m\}$. Since $(\mathbb{A},\alpha)$ is PR-Archimedean, there is a PR-function $f$ with $M_i\leq f(i)$. The second assertion follows from the definition of $\widehat{\alpha}(\langle i,k\rangle)$.
 \qed
 \medskip
  
{\bf Fact 2.} Given a polynomial $p\in\mathbb{A}[x]$ of degree $>1$, one can primitive recursively find the Sturm sequence of polynomials $\sseq(p)=(p_0,p_1,\ldots,p_m)$ in $\mathbb{A}[x]$ with the following property: the number of real roots of $p$ in any interval $ (a,b] $ equals $ v (a) - v (b) $ where $ v (c) $, for $ c \in \real $, is the sign alternation number in the sequence $(p_0(c),p_1(c),\ldots,p_m(c))$.
 \medskip
 
{\em Proof.} As is well known, the sequence $\sseq$ is defined as follows: $p_0=p$, $p_1=p'$ is the derivative of $p$, and for $j>1$, $p_j$ is the negative remainder after dividing $p_{j-1}$ by $p_{j-2}$ (thus, $\sseq(p)$ is a small variation of the sequence from the Euclidean algorithm for $p,p'$). By Proposition \ref{arith} and other remarks in Section \ref{prel}, the sequence $\sseq(p)$ can be found primitive recursively.
 \qed
 \medskip
  
{\bf Fact 3.} Given a non-zero polynomial $p\in\mathbb{A}[x]$ and $a,b\in\rat$, one can primitive recursively find the number of real roots of $p$ in the interval $(a,b]$, as well as  the number of all real roots of $p$.
 \medskip
 
{\em Proof.} Follows from Facts 1, 2 and Proposition \ref{eval}.
 \qed
 \medskip
  
{\bf Fact 4.} Given  $p\in\mathbb{A}[x]$ of degree $m\geq2$, one can primitive recursively find a positive rational $\delta_p<\Delta_p$ where $\Delta_p$  is the smallest distance between distinct roots of $p$.
 \medskip
 
{\em Proof.} Without loss of generality we can think that $p$ has no multiple roots (otherwise, we can take 
$ p  / \gcd (p, p') $ instead of $p$ where $\gcd(p, p')$ is the greatest common divisor of $p$ and $p'$, because $p$ and $ p  / \gcd (p, p') $ are known to have the same roots and the latter polynomial has no multiple roots, see remarks before Proposition \ref{arith}). By Mahler's theorem (see the corollary of Theorem 2 in \cite{mah64}), 
$$\Delta_p > \sqrt{3}m^{-\frac{m+2}{2}}|D(p)|^{{\frac{1}{2}}}|L(p)^{-(m-1)}>m^{-(m+2)}|D(p)|^{{\frac{1}{2}}}|L(p)^{-(m-1)}$$
 where $D(p)$ is the discriminant of $p$  and $L(p)=|\alpha(i_0)|+\ldots+|\alpha(i_m)|$. Since $D(p)\in A$ by Proposition \ref{res} and $p$ has no multiple roots, $D(p)\neq0$. Since $(\mathbb{A},\alpha)$ is PR-Archimedean, we can primitive recursively find a positive rational  $\delta_{p}$ below $m^{-(m+2)}|D(p)|^{{\frac{1}{2}}}|L(p)^{-(m-1)}$.
 \qed
 \medskip
  
{\bf Fact 5.} Given a non-zero polynomial $p\in\mathbb{A}[x]$ and a positive rational number $\varepsilon$, one can primitive recursively find a sequence $I_1<\cdots<I_l$ (where $l\geq0$ is the number of real roots of $p$) of pairwise disjoint rational intervals of length $\leq\varepsilon$ which separate the real roots of $p$, i.e.  every $I_j$ contains  precisely one real root of $p$.
 \medskip

{\em Proof.} Follows from the previous facts using the bisection method. 
 \qed
 \medskip 
 
{\bf Fact 6.} Operations $+,\cdot,-,^{-1}$ on $\widehat{A}$ are  $\widehat{\alpha}$-PR.
 \medskip
 
{\em Proof.} All operations are considered similarly, so we give details only for $+$; we describe a PR function $f:\nat\times\nat\to\nat$ with $\widehat{\alpha}(m)+\widehat{\alpha}(m')=\widehat{\alpha}(f(m,m'))$. Let $m=\langle i,k\rangle$ and $m'=\langle i',k'\rangle$. By the definition of $P(i,k)$, this relation is PR. If $P(i,k)$ then we set $f(m,m')=m'$. If $\neg P(i,k)$ and $P(i',k')$ then we set $f(m,m')=m$. It remains to consider the case when both $P(i,k)$ and $P(i',k')$ are false, i.e. $\widehat{\alpha}(m)=c$ is the $(k+1)$-st real root of $p=\alpha^*_i$ and $\widehat{\alpha}(m')=d$ is the $(k'+1)$-st real root of $q=\alpha^*_{i'}$. It suffices to primitive recursively find $s,t\in\nat$ such that $\neg P(s,t)$ and  $c+d$ is the $(t+1)$-st real root of  $r=\alpha^*_s\in\mathbb{A}[x]$ (then we can set $f(m,m')=\langle s,t\rangle$).

By (the proof of) Theorem 6 in \cite{lo82}, the polynomial $r=\res(p(x-y),q(y))$ has $c+d$ as a root (in fact, the complex roots of $r$ coincide with the sums of complex roots of $p,q$). By remarks before Proposition \ref{res}, we can primitive recursively find $s$ with $r=\alpha^*_s$. For any rational intervals $(a,b)\ni c$ and $(a',b')\ni d$, the interval $I=(a+a',b+b')$ contains $c+d$, and its length may be made arbitrarily small. Using Fact 5, we can primitive recursively find a sequence $I_1<\cdots<I_l$ of rational intervals which separate all real roots of $r$ such that $I$ intersects precisely one interval $I_t$, $t\leq l$, of this sequence. Then $c+d\in I_t$, hence it remains to  set $f(m,m')=\langle s,t\rangle$.
 \qed   
 \medskip
 
{\bf Fact 7.} The relation $\leq$ on $\widehat{A}$ is $\widehat{\alpha}$-PR.
 \medskip
 
{\em Proof.} By Fact 6, it suffices to show that the relation $0\leq\widehat{\alpha}(m)$  is PR. Let again $m=\langle i,k\rangle$. By the  definition of $\widehat{\alpha}(m)$  we have: $0\leq\widehat{\alpha}(m)$ iff either $P(i,k)$ or ($\neg  P(i,k)$ and the $(k+1)$-st real root of $\alpha^*_i$ is non-negative). 
Consider the case when $P(i,k)$ is false. By Fact 5, we  can primitive recursively find a sequence $I_1<\cdots<I_l$ (where $l> k$ is the number of real roots of $\alpha^*_i$) of pairwise disjoint rational intervals of length $\leq\varepsilon$ such that every $I_j$ contains  precisely one real root of $\alpha^*_i$. Then $\widehat{\alpha}(m)\in I_{k+1}$. Assume first that $\alpha^*_i(0)=0$ (i.e., $\alpha(i_0)=0$). Then $0\in I_j$ for a unique $j\leq l$, hence $0\leq\widehat{\alpha}(m)$ iff $j\leq k+1$. 

In the case $\alpha^*_i(0)\not=0$, we consider the polynomial $q=x\alpha^*_i\in\widehat{A}[x]$ which satisfies $q(0)=0$. Computing the sequence $I_1<\cdots<I_l$ for polynomial $q$ in place of $\alpha^*_i$ and applying the argument of the previous paragraph  we see that $0\leq\widehat{\alpha}(m)$ iff $j< k+1$.  
Altogether, these arguments and the primitive recursiveness of relation $P(i,k)$ complete the proof.
 \qed

\medskip
By a classical theorem of Steinitz, for any  field $\mathbb{A}$ there exists its algebraic closure $\overline{\mathbb{A}}\supseteq\mathbb{A}$  which  is unique in the sense that for any algebraic closure $\mathbb{B}$ of $\mathbb{A}$ there is an  isomorphism between $\mathbb{B}$ and $\overline{\mathbb{A}}$ identical on $\mathbb{A}$.
M. Rabin \cite{ra60}  proved a computable version of the Steinitz theorem: if $\mathbb{A}$ is constructivizable then so is also $\overline{\mathbb{A}}$. The uniqueness up to computable isomorphism holds for constructive fields which have computable splitting.
Though   we do not yet know a complete PR-analogue of Rabin's theorem, we note that the algebraic closures of PR Archimedean ordered fields are PR-constructive.

Recall (see e.g. Chapters 10, 11 of \cite{wa67}) that a real closed field  $\mathbb{B}$ is never algebraically closed (the polynomial $x^2+1$ has no root in  $\mathbb{B}$) but its algebraic closure $\overline{\mathbb{B}}$ is constructed very easily by adjoining a root  of $x^2+1$. Thus, $\overline{\mathbb{B}}$ is isomorphic to $\mathbb{B}\times\mathbb{B}$ where the arithmetic on pairs is similar to that of the field $\co$ of complex numbers. If $(\mathbb{A},\alpha)$ is a constructive ordered subfield of $\IR$, let $\bar{\alpha}$ be the induced numbering of $\overline{\mathbb{A}}$ (considered as a subfield of $\co$), i.e. $\bar{\alpha}\langle n_1,n_2\rangle=(\widehat{\alpha}(n_1),\widehat{\alpha}(n_2))$. The following is an immediate corollary of Theorem \ref{main1}.

\begin{Corollary}\label{main21}
If $(\mathbb{A},\alpha)$ is a PR-Archimedean  ordered subfield of $\IR$ then  $(\overline{\mathbb{A}},\overline{\alpha})$ is a PR  subfield of  $\co$.
\end{Corollary}

\begin{Remark}\label{commain}
The computable versions of proofs above show that the computable version of Theorem \ref{main1} holds: If $(\mathbb{A},\alpha)$ is a constructive  ordered subfield of $\IR$ then so is also $(\widehat{\mathbb{A}},\widehat{\alpha})$.
\end{Remark}

\section{PR root-finding}\label{roots}

We say that a computable field $(\mathbb{B},\beta)$ {\em has computable root-finding} (cf. \cite{fs56,mi10}) if, given a polynomial $p\in\mathbb{B}[x]$ of degree $>1$, one can compute a (possibly, empty) list of all roots of $p$ in $\mathbb{B}$. Theorem 4.43 in \cite{fs56} states that  $(\mathbb{B},\beta)$  has computable root-finding iff it has computable splitting. As usual, the notion of PR root-finding is obtained by changing ``computable'' to ``PR'' in the definition above. The proof of Theorem 4.43 in \cite{fs56} works for the PR-version which, together with Proposition \ref{symmetric}, yields the following.

\begin{Proposition}\label{rfinding0}
A PR field  has PR root-finding iff it has PR splitting.
\end{Proposition}

Obviously, every computable algebraically closed field  has computable root-finding, but the  proof makes use of the unbounded search. The next theorem shows that the PR-version of this holds at least for some fields considered above. To shorten  formulations, we denote by $\prr(\IR)$ (resp. $\pr(\IR),\prs(\IR)$) the set of all  $\alpha:\nat\to\IR$ such that $(\mathbb{A},\alpha)$, $A=\rng(\alpha)$, is a PR-constructive ordered subfield of $\IR$ (resp., a PR-Archimedean, a PR-Archimedean ordered subfield with PR splitting). By a PRAS-field we mean an ordered subfield $\mathbb{A}$ if $\IR$ which has a PR-constructivization $\alpha\in\prs(\IR)$. 
In the remaining part of the paper, we usually denote by  $\mathbb{A},\widehat{\mathbb{A}},\overline{\mathbb{A}}$ the (ordered) fields associated with $\alpha,\widehat{\alpha},\overline{\alpha}$, respectively. 

\begin{Theorem}\label{rfinding}
If $\alpha\in\prs(\IR)$ then  $(\widehat{\mathbb{A}},\widehat{\alpha})$ and $(\overline{\mathbb{A}},\overline{\alpha})$  have PR root-finding.
\end{Theorem}

Under the assumption $\alpha\in\prs(\IR)$ we first establish some auxiliary facts. Let $\sigma=\{+,\cdot,c_0,c_1,\ldots\}$.

 \medskip

{\bf Fact 1.} Given  a polynomial $p\in \mathbb{A}[x_0,\ldots,x_k]$ and  polynomials $p_0,\ldots,p_k\in\mathbb{A}[x]$ of positive degrees, one can primitive recursively find $q\in\mathbb{A}[x]$ of positive degree such that, for all complex roots $b_i$ of $p_i$, $i\leq k$, the value $p(b_0,\ldots,b_k)$ is a root of $q$.
 \medskip
 
{\em Proof.} Let $r^+,r^\cdot$ be  binary operators on $\mathbb{A}[x]$ such that all sums (resp. products) of complex roots of $p,q$ are among the  roots of $r^+(p,q)$ (resp. $r^\cdot(p,q)$).  By the proof of Theorem 6 in \cite{lo82} (see also the proof of Fact 6 in the previous section), operators  $r^+,r^\cdot$ are given by explicit formulas (based on a resultant calculus) which show they are $\alpha^*$-PR.

We associate with any $\sigma$-term $t=t(x_0,\ldots,x_k)$ the polynomial $q_t\in\mathbb{A}[x]$ as follows: $q_{c_n}=x-\alpha(n)$, $q_{x_i}=p_i$, $q_{t_1+t_2}=r^+(q_{t_1},q_{t_2})$, $q_{t_1\cdot t_2}=r^\cdot(q_{t_1},q_{t_2})$. By induction on $t$ one easily checks that, for all complex roots $b_i$ of $p_i$,  $i\leq k$, $t^\mathbb{A}(b_0,\ldots,b_k)$ is a root of  $q_t$.
  
By Proposition \ref{polterm}, from   $p\in\mathbb{A}[x_0,\ldots,x_k]$ one can primitive recursively find a $\sigma$-term $t=t(x_0,\ldots,x_k)$ such that $p(b_0,\ldots,b_k)=t^\mathbb{A}(b_0,\ldots,b_k)$ for all $b_0,\ldots,b_k\in A$. Thus, we can take $q=q_t$.
 \qed
 \medskip
  
{\bf Fact 2.}  Given a non-zero polynomial  $r\in\mathbb{A}[x]$, one can primitive recursively find $q_1,q_2\in\mathbb{A}[x]$ such that, for any complex root $b=(b_1,b_2)$ of $r$, the real part $b_1$ is a root of $q_1$ and the imaginary part $b_2$ is a root of $q_2$.
 \medskip
 
{\em Proof.} Let $q_1=q$ be the polynomial obtained from the algorithm of Fact 1 for $ p (x_{0},x_{1}) = \frac{1}{2}(x_{0}+x_{1}) $ and $p_0=p_1=r$. For any complex root $b=(b_1,b_2)$ of $r$ we then  have: $p(b,\bar{b})=b_1$, $b$ is a root of $p_0$, and $\bar{b}$ is a root of $p_1$ where $\bar{b}=(b_1,-b_2)=b_1-ib_2$ is the complex conjugate of $b$ and $i$ is the imaginary unit. Thus, $q_1$ has the desired property. 

Let now $q=a_0+a_1x+a_2x^2+\cdots$ be the polynomial obtained from the algorithm of Fact 1 for $ p (x_{0},x_{1}) = \frac{1}{2}(x_{0}-x_{1}) $ and $p_0=p_1=r$. For any complex root $b=(b_1,b_2)$ of $r$ we then  have: $p(b,\bar{b})=ib_2$, $b$ is a root of $p_0$, and $\bar{b}$ is a root of $p_1$. Thus, $q(ib_2)=0$, hence also $q(-ib_2)=0$. Summing up the last two equalities we see that $q_2(b_2)=0$ where $q_2=a_0+a_2x^2+\cdots$. Thus, $q_2$ has the desired property. 
 \qed
 \medskip
  
 
  
{\bf Fact 3.} Given a  polynomial $p=b_0x^0+\cdots+b_nx^n\in\overline{\mathbb{A}}[x]$ and polynomials $q_0,\ldots,q_n\in\mathbb{A}[x]$ such that $q_i(b_i)=0$ for each $i\leq n$, one can primitive recursively find $r\in\mathbb{A}[x]$ such that all complex  roots of $p$ are among the roots of $r$.
 \medskip
 
{\em Proof.} The proof is essentially the same as in the Algorithm 3 of \cite{lo82}, hence we give only a sketch. Since $(\mathbb{A},\alpha)$ has PR splitting, we may without loss of generality think that $q_0,\ldots,q_n$ are irreducible. By Proposition \ref{prmitive}(3), we may primitive recursively find $b\in\overline{\mathbb{A}}$ and irreducible $t,p_0,\ldots,p_n\in\mathbb{A}[y]$ such that $\mathbb{A}(b)=\mathbb{A}(b_0,\ldots,b_n)$ $t(b)=0$, and $p_i(b)=b_i$ for all $i\leq n$. Let $s=\gcd(t,p_n)$. We may without loss of generality think that $\deg(s)=0$ (otherwise, replace $t$ by $t/s$). Then the polynomial $r=\res(t,q)$, where $q=p_0x^0+\cdots+p_nx^n\in\mathbb{A}[y][x]$, has the desired properties. By the results in Section \ref{prings}, $r$ may be found primitive recursively.
 \qed
 \medskip
  
{\em Proof of Theorem \ref{rfinding}.} Given $p\in\overline{\mathbb{A}}[x]$, we have to primitive recursively find all complex roots of $p$. By Fact 3, we can find  $r\in\mathbb{A}[x]$ such that all complex  roots of $p$ are among the roots of $r$. By Fact 2, we can find $q_1,q_2\in\mathbb{A}[x]$ such that, for any complex root $b=(b_1,b_2)$ of $r$, the real part $b_1$ is a root of $q_1$ and the imaginary part $b_2$ is a root of $q_2$. By the definition of $\overline{\alpha}$ and the results of Section \ref{madis}, we can find the lists $b_{1,0}<\cdots<b_{1,m}$ and $b_{2,0}<\cdots<b_{2,n}$ of all real roots of $q_1$ and $q_2$, respectively. Then all complex roots of $p$ are among $(b_{1,i},b_{2,j})$ where $i\leq m,j\leq n$. Substituting these complex numbers one by one in $p$ and using Proposition \ref{eval}, we can primitive recursively choose all  complex roots of $p$.

It remains to show that $(\widehat{\mathbb{A}},\widehat{\alpha})$  has PR root-finding. Let $p\in\widehat{\mathbb{A}}[x]$; we have to find a list of all real roots of $p$. Since $\widehat{\alpha}\leq_{PR}\overline{\alpha}$, we have $\widehat{\alpha}^*\leq_{PR}\overline{\alpha}^*$ by Proposition \ref{polynom}. By the previous paragraph, we can compute the list of all complex roots of $p$. Choosing the real numbers from this list, we obtain a list of all real roots of $p$.
 \qed
 \medskip 

\begin{Corollary}\label{rfinding1}
\begin{enumerate}\itemsep-1mm
 \item For all $\alpha,\beta\in \pr(\IR)$, we have: $\alpha\leq_{PR}\widehat{\alpha}\in\pr(\IR)$,  $\alpha\leq_{PR}\beta$ implies $\widehat{\alpha}\leq_{PR}\widehat{\beta}$, and similarly for $\prs(\IR)$ in place of $\pr(\IR)$.
 \item  For any $\alpha\in \prs(\IR)$, $\widehat{\widehat{\alpha}}\leq_{PR}\widehat{\alpha}\in\prs(\IR)$, so $\alpha\mapsto\widehat{\alpha}$ is a closure operator on $(\prs(\IR);\leq_{PR})$.
\end{enumerate}
\end{Corollary}

As mentioned in Section \ref{madis}, the real closure of an ordered field is unique up to an isomorphism over this ordered field. We conclude this section with a PR-version of this fact. The proof is an easy adaptation of the proof for the classical case (see e.g. \cite{wa67}, Theorem 8 in Section 82).

\begin{Proposition}\label{rfinding2}
Let  $\alpha\in \prs(\IR)$ and let $(\mathbb{B},\beta)$ be a PR-Archimedean ordered field such that it has PR root-finding, $\mathbb{B}$ is a real closure of $\mathbb{A}$ and $\alpha\leq_{PR}\beta$. Then $(\widehat{\mathbb{A}},\widehat{\alpha})$ is PR isomorphic to $(\mathbb{B},\beta)$ over $(\mathbb{A},\alpha)$, i.e. there is an (unique) isomorphism $f$ from $\widehat{\mathbb{A}}$ onto $\mathbb{B}$ such that $f$ is identical on $A$, $f$ is $(\widehat{\alpha},\beta)$-PR, and $f^{-1}$ is $(\beta,\widehat{\alpha})$-PR.
\end{Proposition}

\begin{Remark}\label{comunique}
Let $\com(\IR), \cs(\IR)$ be the computable analogues of $\pr(\IR), \prs(\IR)$, i.e. $\com(\IR)$ (resp. $\cs(\IR)$) is the set of all maps $\alpha:\nat\to\IR$ such that $(\mathbb{A},\alpha)$ is a constructive ordered subfield of $\IR$ (resp., a constructive  ordered subfield which has computable  splitting). 
The proofs above apply to these computable analogues,  in particular the computable version of Proposition \ref{rfinding2} gives a sufficient condition for uniqueness of the computable real closure. 
\end{Remark}

\section{PRAS-Fields vs. PR reals}\label{prreals}

In \cite{ss17,ss17a} we have shown that for any finite set $F$ of computable reals there is a computable real closed ordered subfield $(\mathbb{B},\beta)$ of the computable reals such that $F\subseteq B$ (see also a more general Theorem 4.1 in \cite{mg19} obtained independently). This fact implies the following charaterization of the union of all constructivizable ordered fields of reals:
 $$\rc=\bigcup\{A\mid \alpha\in \com(\IR)\}=\bigcup\{A\mid \alpha\in \cs(\IR)\}$$
  where $\rc$ is the set of computable reals and $\com(\IR), \cs(\IR)$ are the computable analogues of $\pr(\IR), \prs(\IR)$.
   In this section  we search for  PR-analogues of these facts from \cite{ss17,ss17a}.

The PR-analogue of $\rc$ is the ordered field $\rp$ of PR reals. In this section we often cite the survey \cite{zh07} on PR reals where references to the source papers may be found; we sometimes use slightly different notation from that in \cite{zh07}. Recall that a real number $a$ is PR if $a=\lim_nq_n$ for a PR sequence $\{q_n\}$ of rational numbers which is fast Cauchy, i.e. $|q_n-q_{n+1}|<2^{-n}$ for all $n$. There is a natural  numbering $\tau$ of $\rp$. To define it precisely, we describe some technical details which will also be used in Section \ref{anal}.  

With any $p\in\calN$ (see the end of Section \ref{prel}) we associate the sequence $\varkappa\circ p\in\rat^\nat$ of rational numbers. Since $\varkappa:\nat\to\rat$ is a bijection, the function $p\mapsto\varkappa\circ p$ is a bijection between $\calN$ and $\rat^\nat$. From the properties of $\varkappa$ in Section \ref{prel} it follows that  $p$  is PR iff $\varkappa\circ p$ is PR, and similarly for functions of larger arity. By $\cau$ we denote the set of all $p\in\calN$ such that $\varkappa\circ p$ is fast Cauchy. With any $p\in\calN$ we associate $\tilde{p}\in\calN$ as follows: if $\forall i<n(|\varkappa(p(i))-\varkappa(p(i+1))|<2^{-i})$ then $\tilde{p}(n)=p(n)$, else $\tilde{p}(n)=p(i_0)$ where $i_0=\mu i<n(|\varkappa(p(i))-\varkappa(p(i+1))|\geq2^{-i})$. Clearly, $\tilde{p}\in\cau$ for each $p\in\calN$, and if $p\in\cau$ then $\tilde{p}=p$, in particular $\cau=\{\tilde{p}\mid p\in\calN\}$.

Now we define the numbering $\tau$ of $\rp$ by: $\tau(e)=\lim_n\varkappa(\tilde{\psi}_e(n))$ where $\psi$ is the numbering of unary PR functions from the end of Section \ref{prel} and $\psi_e=\psi(e)$. In the next proposition we collect some properties of the introduced objects. 

\begin{Proposition}\label{pi}
\begin{enumerate}\itemsep-1mm
 \item $\rp$ is a subfield of $\mathbb{R}$, and operations $+,-,\cdot$ are $\tau$-PR.
 \item The sequence of reals $\{\tau(e)\}$ is computable, hence the inclusion $\rp\subseteq\rc$ is proper.
 \item The relation $<$ on $\rp$ is $\tau$-c.e.
 \item The ordered field $\rp$ is real closed.
 \item The ordered field $\rp$ is not constructivizable.
\end{enumerate}
 \end{Proposition}
 
 {\em Proof.} (1) Proof is easily extracted (say) from the proofs of Proposition 4.1 and Theorem 4.2 in \cite{zh07} and the remarks at the end of Section \ref{prel}. 
We have to find a PR function $p$ such that $\tau(m)+\tau(n)=\tau(g(m,n))$, and similarly for $-,\cdot$ is simpler). Denote the sequences $\tilde{\psi}_m$ and $\tilde{\psi}_n$ by $\{p_i\}$ and $\{q_i\}$, respectively. These are PR fast Cauchy sequences of rationals converging to $\tau(m)$ and $\tau(n)$, respectively.

Then $|p_{i+1}-p_{j+1}|<2^{-k}$ and $|q_{i+1}-q_{j+1}|<2^{-k}$ for $i,j\geq k$. Let $r_i=p_i+q_i$, then $\{r_{i+2}\}$ is a fast Cauchy sequence of rationals that converges to $\tau(m)+\tau(n)$. By the properties of $\varkappa$, $h=\{\varkappa^{-1}(r_{i+2})\}$, $h=h_{m,n}$ is a PR function from $\cau$, hence $\tilde{h}=h$. By Robinson's theorem, $h_{m,n}=\bft_{m,n}$ for some $t_{m,n}\in T_0$. Moreover, the terms $t_{m,n}$ may be chosen so that the function $(m,n)\mapsto N(t_{m,n})$ is PR. Then for some binary PR function $g$ we have $h_{m,n}=\psi_{g(m,n)}$. Therefore, $\tau(m)+\tau(n)=\tau(g(m,n))$, as desired.

For the operation $-$ the proof is the same, only now we have to set $r_i=p_i-q_i$. For the operation $\cdot$ the proof is the same, only now we have to set $r_i=p_i\cdot q_i$ and $h=\{\varkappa^{-1}(r_{i+f(m,n,i)})\}$, for a suitable PR function $f$ (see the proof of Theorem 4.2 in \cite{zh07}). 
 
 (2) The  computability of the sequence follows from the definition of $\tau$. The properness of inclusion follows by the well-known property of the computable reals.
 
 (3) Obvious from the definition of $\tau$.
 
 (4) This is due to Peter Hertling (private communication \cite{her20}, with permission to mention the formulation here). 
 
 (5) This is proved in \cite{kh87}.
 \qed 

\begin{Remark}\label{pi0}
 The reciprocal partial function $x^{-1}$ is not $\tau$-PR (this is proved by essentially the same argument as that in Proposition \ref{samples}(2); the argument is topological, hence more natural to be discussed in Section \ref{anal}). Nevertheless, the restriction of the reciprocal  function to $\rp\setminus(-\frac{1}{n+1},\frac{1}{n+1})$ is $\tau$-PR uniformly on $n$.
 \end{Remark}

For any $\alpha\in \pr(\IR)$, let $\rp(\alpha)$ be the set of PR reals $b$ such that the sign of polynomials in $\mathbb{A}[x]$ at $b$ is checked primitive recursively. Formally, for any real $b$, let $\sign(b)$ be $0,1,2$ depending on whether $b$ is  zero,  positive, or negative. Then $\rp(\alpha)$ is the set of all $b\in\rp$ such that the function $i\mapsto\sign(\alpha_i^*(b))$ is PR. More generally, for any $n\geq0$, let $\rp^{[n]}(\alpha)$ be the set of tuples $\bar{b}=(b_0,\ldots,b_n)$ of PR reals such that the function $i\mapsto\sign(\alpha_i^{[n]}(b_0,\ldots,b_n))$ is PR where $\alpha^{[n]}$ is the numbering of $\mathbb{A}[x_0,\ldots,x_n]$ from Section \ref{polynom}. Note that $\rp(\alpha)=\rp^{[0]}(\alpha)$.

\begin{Proposition}\label{rpalpha}
\begin{enumerate}\itemsep-1mm
 \item If $\alpha,\beta\in \pr(\IR)$ and $\alpha\leq_{PR}\beta$ then $\rp(\beta)^{[n]}\sq\rp(\alpha)^{[n]}$ for all $n$.
 \item For all $\alpha\in \pr(\IR)$ and $n$ we have: $\rp(\alpha)^{[n]}\sq\rp(\varkappa)^{[n]}$.
 \item For any $\alpha\in \prr(\IR)$ we have   $A\sq\rp(\alpha)$.
 \item For any $\alpha\in \pr(\IR)$ there is  a uniformly PR sequence $\{g_n\}$ of PR functions $g_n:\nat\to\rat$ such that $\{g_n(i)\}_i$ is a fast Cauchy sequence converging to $\alpha(n)$, in particular $\alpha\leq_{PR}\tau$.
 \end{enumerate}
 \end{Proposition} 

{\em Proof.} 1. Let $\bar{b}\in\rp(\beta)$, so $i\mapsto\sign(\beta_i^*(\bar{b}))$ is PR. Since $\alpha\leq_{PR}\beta$,  by Proposition \ref{polynom} we have $\alpha^{[n]}\leq_{PR}\beta^{[n]}$, so $\alpha^{[n]}_i=\beta^{[n]}_{f(i)}$ for some PR function $f$. Then $\sign(\alpha_i^{[n]}(\bar{b}))=\sign(\beta^{[n]}_{f(i)}(\bar{b}))$, hence  $i\mapsto\sign(\alpha_i^{[n]}(\bar{b}))$ is PR.

2. The assertion follow from item 1 because, clearly, $\varkappa\leq_{PR}\alpha$.

3. Let $a\in A$ and let $a_0,a_1\in\intt$ satisfy $a_0<a<a_1$. Using the bisection method starting with the interval $[a_0,a_1]$ and the PR-constructivity of $\alpha$, we construct a PR function $g:\nat\to\rat$ such that   $|g(i)-a|\leq(a_1-a_0)\cdot 2^{-i}$. Then $\{g(i)\}_i$ is a fast Cauchy sequence converging to $a$, hence $a\in\rp$. Moreover, by Proposition \ref{polynom} we get $a\in\rp(\alpha)$.

4. Since $(\mathbb{A},\alpha)$ is PR-Archimedean, $-f(n)<\alpha(n)<f(n)$ for some PR function $f$. Using the bisection method as in (3) with $[-f(n),f(n)]$ in place of $[a_0,a_1]$, we construct a uniformly PR sequence $\{g_n\}$ with the specified properties. The uniformity yields a PR function $f$ such that $\varkappa^{-1}\circ g_n=\psi_{f(n)}$. By the definition of $\tau$, $\alpha\leq_{PR}\tau$ via $f$. 
 \qed

\medskip
By items 2 and 3 of the last proposition, $\alpha\in \pr(\IR)$ implies $\rng(\alpha)\sq\rp(\varkappa)$, hence all PR ordered fields of reals are contained in $\rp(\varkappa)$. The next result shows which elements of $\rp(\varkappa)$ (and, moreover, which finite tuples of reals) can be included into some  PR ordered field of reals. 

\begin{Theorem}\label{ext}
 Let $\alpha\in \prs(\IR)$ and $\bar{b}\in\rp^n$. Then $\exists\beta\in \prs(\IR)(\alpha\leq_{PR}\beta\wedge b_0,\ldots,b_n\in B)$ iff $\bar{b}\in\rp^{[n]}(\alpha)$.
\end{Theorem}

{\em Proof.} If $\beta\in \pr(\IR)$, $\alpha\leq_{PR}\beta$, and $b_0,\ldots,b_n\in B=\rng(\beta)$, then $i\mapsto\sign(\alpha_i^{[n]}(\bar{b}))$ is PR by Proposition \ref{polynom}, hence $\bar{b}\in\rp^{[n]}(\alpha)$. 

Conversely, let $\bar{b}\in\rp^{[n]}(\alpha)$. 
First we consider the case when $b_0,\ldots,b_n$ are algebraically independent over $\mathbb{A}$, hence $\sign(\alpha_i^{[n]}(\bar{b}))\in\{1,2\}$ for every $i$ with $\alpha_i^{[n]}\neq0$. Then $\mathbb{A}(b_0,\ldots,b_n)$ with the induced numbering $\gamma$ is a PR field by Proposition \ref{rfunc} because it is isomorphic to the field $\mathbb{A}(x_0,\ldots,x_n)$ of rational functions. The elements of $\mathbb{A}(b_0,\ldots,b_n)$ have the form $p(\bar{b})/q(\bar{b})$ for some $p,q\in\mathbb{A}[x_0,\ldots,x_n]$, $q\neq0$. Since $p(\bar{b})/q(\bar{b})>0$ iff both $p(\bar{b}),q(\bar{b})$ are positive or both $p(\bar{b}),q(\bar{b})$ are negative, $\gamma$ is a PR constructivization of $\mathbb{A}(b_0,\ldots,b_n)$. By Proposition \ref{prsplit}, $\gamma\in \prs(\IR)$, hence we can take $\beta=\gamma$.

Now let $\bar{b}\in\rp^n$ be arbitrary. Without loss of generality (renumbering $b_0,\ldots,b_n$ if necessary), let $j\leq n$ be the unique number such that $b_0,\ldots,b_{j-1}$ are algebraically independent over $\mathbb{A}$ while $b_j,\ldots,b_{n}$ are algebraic over $\mathbb{A}(b_0,\ldots,b_{j-1})$. Let $\gamma$ be defined as in the previous paragraph for $j>0$ (with $n$ replaced by $j-1$) and $\gamma=\alpha$ for $j=0$. By the previous paragraph and Proposition \ref{rfinding1}, we can take $\beta=\widehat{\gamma}$.
 \qed

\medskip
The results above imply the following characterization of the union of all PRAS-fields.

\begin{Corollary}\label{rpalpha1}
 $\rp(\varkappa)=\bigcup\{A\mid \alpha\in \prr(\IR)\}=\bigcup\{A\mid \alpha\in \pr(\IR)\}=\bigcup\{A\mid \alpha\in \prs(\IR)\}$.
\end{Corollary}

{\em Proof.} Obviously, $\bigcup\{A\mid \alpha\in \prr(\IR)\}\supseteq\bigcup\{A\mid \alpha\in \pr(\IR)\} \supseteq\bigcup\{A\mid \alpha\in \prs(\IR)\}$. By Proposition \ref{rpalpha}(3), $\rp(\varkappa)\supseteq\bigcup\{A\mid \alpha\in \prr(\IR)\}$, so it suffices to show that $\rp(\varkappa)\subseteq\bigcup\{A\mid \alpha\in \prs(\IR)\}$. This follows from Theorem \ref{ext} because $\varkappa\in\prs(\IR)$.
 \qed

\medskip
This corollary  would yield  satisfactory PR-analogues of some facts mentioned in the beginning of this section, if we had $\rp(\varkappa)=\rp$. In the next proposition we will show that this is not the case. 

First we recall some facts about  continuous fractions. Associate with any real $x$ the sequence of integers $\{x_i\}$ (which we call here the canonical continuous fraction for $x$) and an auxiliary sequence $\{d_i\}$ as follows: 

let $x_0=\lfloor x\rfloor$ be the integer part of $x$ and $d_0:=x-\lfloor x\rfloor$; 

if $d_0=0$ then let $x_1=0$, otherwise let  $x_1=\lfloor d_0^{-1} \rfloor$ and $d_1:=d_0^{-1}-\lfloor d_0^{-1} \rfloor$;

if $x_{i+1}=0$ or $d_{i+1}=0$ then let $x_{i+2}=0$, otherwise let $x_{i+2}=\lfloor d_{i+1}^{-1} \rfloor$ and $d_{i+2}:=d_{i+1}^{-1}-\lfloor d_{i+1}^{-1} \rfloor$.

 It is well known that $x$ is rational iff $\exists i(x_{i+1}=0)$. It is easy to see that $0\leq d_i<1$, $x_{i+1}\geq0$, $x_{i+1}=0$ implies $x_{i+2}=0$, and $x_{i+1}>0\wedge x_{i+2}=0$ imply $x_{i+1}\geq2$. Note that the set of all integer sequences $\{x_i\}$ with these properties is in a bijective correspondence with $\IR$ which is given by $\{x_i\}\mapsto x=\lim_ny_n$ where 
 $y_0=x_0$, $y_1=x_0+\frac{1}{x_1}$ for $x_1\not=0$ and $y_1=y_0$ otherwise, $y_2=x_0+\frac{1}{x_1+\frac{1}{x_2}}$ for $x_2\not=0$ and $y_2=y_1$ otherwise, and so on. The converse map  $x\mapsto\{x_i\}$ is given by ``computing'' the  canonical continuous fraction for $x$.
 
 The set of reals with the PR  canonical continuous fraction has a natural numbering $\gamma$ constructed as follows. Associate with any sequence $\nu=\{x_i\}$ of integers the sequence $\nu'=\{x'_i\}$ as follows: $x'_0=x_0$; if $x_1<1$ then $x'_1=0$, otherwise $x'_1=x_1$; if $x_{i+1}=0$ or $x_{i+2}<1$ then $x'_{i+2}=0$, otherwise $x'_{i+2}=x_{i+2}$. Then $\nu'$ always has the properties above, and if $\nu$ has these properties then $\nu'=\nu$. Let now $\{\nu_n\}$ be the  sequence of PR integer sequences induced by the computable sequence $\{\psi_n\}$ of all unary PR functions (as usual, we implicitly use a suitable PR bijection between $\nat$ and $\intt$); then the sequence $\{\nu'_n\}$ consists of  all PR sequences with the properties above. Setting $\gamma(n)=\lim\nu'_n$, we obtain a natural numbering $\gamma:\nat\to\IR$ of all reals with the PR  canonical continuous fraction.

\begin{Proposition}\label{inclus}
The inclusion $\rp(\varkappa)\subseteq\rp$ is strict.
\end{Proposition}

{\em Proof.} Let $\IR_4$ be the set of reals which have  PR continuous fractions (see  \cite{zh07}) and references therein).
 By a theorem of Lehman, the inclusion $\IR_4\sq\rp$ is strict (for a proof see e.g. Theorem 7.4 in \cite{zh07}). So, it suffices to show that $\rp(\varkappa)\subseteq\IR_4$. By Corollary \ref{rpalpha1}, it suffices to show that $A\sq\IR_4$ for every $\alpha\in\pr(\IR)$.
 
First we show that  $n\mapsto\lfloor\alpha_n\rfloor$ is a PR  function from $\nat$ to $\intt$. Let $h:\intt\to\nat$ be a PR function such that $z=\alpha_{h(z)}$ for every $z\in\intt$. Since $\alpha\in\pr(\IR)$, for some PR function $f$ we have $-f(n)<\alpha_n<f(n)$, hence $0<\alpha_n+f(n)<2f(n)$. Since $\lfloor\alpha_n\rfloor=\lfloor\alpha_n+f(n)\rfloor-f(n)$, it suffices to show that the function $n\mapsto\lfloor\alpha_n+f(n)\rfloor$ is PR. But this follows from PR-constructivity of $(\mathbb{A},\alpha)$ and the obvious equalities
\begin{eqnarray*}
 \lfloor\alpha_n+f(n)\rfloor=\mu m<2f(n)(m\leq\alpha_n+f(n)<m+1)\\=\mu m<2f(n)(\alpha_{h(m)}\leq\alpha_n+\alpha_{h(f(n)))}<\alpha_{h(m+1)}).
 \end{eqnarray*} 
 Let $\{x^n_i\}_i$ be the canonical continuous fraction for $\alpha_n$. By the definition  above, it is a PR sequence of rationals (even uniformly on $n$), hence $\alpha_n\in\IR_4$ and therefore $A\sq\IR_4$.
 \qed

We guess that the inclusion $\rp(\varkappa)\sq\IR_4$ is also strict. 

\begin{Remarks} 1. From the  proof it follows that for any $\alpha\in\pr(\IR)$ the relation $\alpha_n\in\intt$ is PR (because it is equivalent to $\alpha_n=\alpha_{h(\lfloor\alpha_n\rfloor)}$).

2. Also, the proof shows that $\alpha\leq_{PR}\gamma$ (this follows from the definition of  the numbering $\gamma:\nat\to\IR$ above and the fact that the sequence $\{x^n_i\}_i$ is PR uniformly on $n$).

3. For the computable versions of $\pr(\IR)$ and $\prs(\IR)$ all proofs of this section  remain valid and yield the results mentioned in the beginning of this section because, as one easily checks, $\rc(\varkappa)=\rc$ and the function $i\mapsto\sign(\alpha_i^{[n]}(\bar{b}))$ is computable for every $\bar{b}\in\rc$. Note that the computable versions of $\prr(\IR)$ and $\pr(\IR)$ coincide.
\end{Remarks}

We see that the relations between classes $\{\mathbb{A}\mid \alpha\in \prr(\IR)\},\;\{\mathbb{A}\mid \alpha\in \pr(\IR)\},\;\{\mathbb{A}\mid \alpha\in \prs(\IR)\}$ and fields $\rp$, $\IR_4$ are much more intricate than for their computable analogues.

How rich is the collection $\prs(\IR)$? By Proposition \ref{prsplit}(1), $\varkappa\in\prs(\IR)$. By Proposition \ref{rfinding1}, $\prs(\IR)$ is closed under $\alpha\mapsto\widehat{\alpha}$, hence $\widehat{\varkappa}\in\prs(\IR)$, i.e., $\ra$ is a PRAS-field. In fact, $\ra$ is even PTIME-presentable \cite{as18,as19}. Moreover, it is open whether there is PTIME-presentable ordered field of reals beyond $\ra$. In the next section we show that PRAS-fields may contain transcendental numbers.

\section{PRAS-Fields beyond $\ra$}\label{prtr}

In this section we examine which transcendental PR reals (and tuples of PR reals) may be included in a PRAS-field. First we show that there are  PRAS-fields of any countable transcendence degree.

\begin{Theorem}\label{trans}
\begin{enumerate}\itemsep-1mm
 \item For any $\alpha\in \prs(\IR)$ and any non-empty rational interval $I$ there exists $b\in I\cap\rp(\alpha)$ which is transcendental over $\mathbb{A}$.
 \item For any $\alpha\in \prs(\IR)$ there exists a uniformly PR infinite sequence $b_0,b_1,\ldots$ of reals which are algebraically independent over $\mathbb{A}$ and satisfy $\bar{b}\in\rp^{[n]}(\alpha)$ for every $n$ where $\bar{b}=(b_0,\ldots,b_{n-1})$.
\end{enumerate}
\end{Theorem}

{\em Proof.} 1. We define by induction PR sequences $\{q_j\}$ of rational numbers and $\{I_j\}$ of rational open intervals such that $q_0\in I_0= I$ and, for every $j$, $I_j\supseteq [I_{j+1}]$ where $[I_{j+1}]$ is the closure of $I_{j+1}$, and $q_{j+1}\in I_{j+1}\sq(q_j-2^{-j},q_j+2^{-j})$. Then we set $b=\lim_jq_j$ which automatically guarantees that $\{q_j\}$ is fast Cauchy and hence $b\in\rp$. The remaining properties of $b$ are obtained by taking some additional care.

Let $I_0=I$ and $q_0$ be any rational number in $I_0$. Assume by induction that we already have defined $q_j,I_j$ for $j\leq n$ which satisfy the properties above for  $j< n$. Then we define $q_{n+1},I_{n+1}$ as follows. If the polynomial $\alpha^*_n$ is zero or has no real roots, choose $q_{n+1},I_{n+1}$ arbitrarily such that $I_n\supseteq [I_{n+1}]$  and $q_{n+1}\in I_{n+1}\sq(q_{n}-2^{-n},q_{n}+2^{-n})$. Otherwise, use Fact 5 in Section \ref{madis} to primitive recursively find a non-empty rational open interval $J$ such that $[J]\subseteq I_n$ and $[J]$ contains no real root of $\alpha^*_n$; then $\alpha^*_n$ is either positive on $[J]$ or negative on $[J]$, and this alternative is checked primitive recursively. Now choose $q_{n+1},I_{n+1}$ as above but with the  additional property $I_{n+1}\subseteq J$. Then the sequences $\{q_j\}$, $\{I_j\}$ are PR and satisfy the properties specified in the previous paragraph.

Note that $b\in I_n$ for all $n$, and if $\alpha^*_n$ is non-zero then it is either positive or negative on $I_{n+1}\ni b$, hence $\alpha^*_n(b)\neq0$; therefore, $b$ is transcendental over $\mathbb{A}$.
If $\alpha^*_n$ is zero then $\sign(\alpha^*_n(b))=0$, otherwise $\sign(\alpha^*_n(b))=1,2$ depending on whether  $\alpha^*_n$ is  positive on $[J]$ or is negative on $[J]$. Therefore, the function $n\mapsto\sign(\alpha^*_n(b))$ is PR and hence $b\in\rp(\alpha)$.

2. Note that the construction $\alpha\mapsto b$ in item 1 is PR in the sense that, given an index for $\alpha\in \prs(\IR)$ (i.e., a code of tuple of indices for PR functions representing the equality and the signature symbols, and also of the splitting function), one can primitive recursively find a $\pi$-index for $b$ and a PR-index of  the function $n\mapsto\sign(\alpha^*_n(b))$.

Let $b_0=b$. Since the construction in Proposition \ref{ext} is PR, we can primitive recursively find an index of $\beta\in \prs(\IR)$ with $\rng(\beta)=\mathbb{A}(b)$. Since the construction in Theorem \ref{madis} is PR, we can primitive recursively find an index of $\widehat{\beta}\in \prs(\IR)$ with $\rng(\widehat{\beta})=\widehat{\mathbb{A}(b)}$. 

Taking $\widehat{\beta}$ in place of $\alpha$, we primitive recursively find an index of some $b_1$ transcendental over $\widehat{\mathbb{A}(b_0)}$. It is easy to check that $b_0,b_1$ are algebraically independent over $\mathbb{A}$ and $(b_0,b_1)\in\rp^{[2]}(\widehat{\beta}))$. Iterating this process indefinitely, we obtain a desired sequence $b_0,b_1,\ldots$.
 \qed
 
\begin{Corollary}
There are  PRAS-fields of any countable transcendence degree.
\end{Corollary}

{\em Proof.}  By Theorems \ref{trans}(2) (for $\alpha=\varkappa$) and \ref{ext}, any of the ordered fields $\rat$, $\rat(b_0)$, $\rat(b_0,b_1),\ldots$, $\rat(b_0,b_1,\ldots) $ is a PRAS-field.  
 \qed

\begin{Remarks}
1. Theorem \ref{trans} and Proposition \ref{ext} imply that the algebraic closures of the fields $\rat(x_0,\ldots,x_n)$ for every $n$, and of $\rat(x_0,x_1,\ldots)$, are PR-constructivizable (the latter fact is a step to proving the PR-version of the Rabin theorem mentioned in Section \ref{madis}).

2. From the construction of numbers $b,b_0,b_1,\ldots$ in the proof Theorem \ref{trans} it follows that all these numbers are in the set PRT defined below. 
\end{Remarks}

An important problem is to determine whether a given concrete transcendental real (or a tuple of such reals) may be included in some PRAS-field. In general, this   question seems very difficult but for some concrete numbers (including the Euler number $e$ and  the circle number $\pi$) the solution follows from results of Goodstein's monograph  \cite{good61}.

By a PR-Cauchy sequence we mean a PR sequence $\{s_i\}$ of rationals such that for some PR function $N$ it holds: $\forall k\forall m,n\geq N(k)(|s_m-s_n|<2^{-k})$. Since further results of this section essentially depend on notions and results of the appendix to \cite{good61}, we make some of our notations closer to those in \cite{good61}. Let $\{P_r\}_{r\geq0}$ be the natural numbering of non-zero rational polynomials defined by $P_r=\varkappa^*_{f(r)}$ where $\varkappa^*$ is the numbering of $\rat(x)$ from Section \ref{prings} and $f$ is the PR function enumerating the PR set $n\mid\varkappa^*_n\not=0$ in the increasing order.

\begin{Definition}\cite{good61}\label{prtrans}
By a PR transcendental real we mean a PR-Cauchy sequence  $\{s_i\}$ such that for some PR functions $k,N$ we have: $|P_r(s_n)|>2^{k(r)}$ for every $n\geq N(r)$. Let PRT denote the set of limits of PR transcendental reals. 
\end{Definition}

\begin{Lemma}\label{prtrans1}
\begin{enumerate}\itemsep-1mm
 \item The set of limits of PR-Cauchy sequences coincides with $\rp$.
 \item If $\{s_i\}$  and $\{q_i\}$ are PR-Cauchy sequences then so are also $\{s_i+q_i\}$, $\{s_i-q_i\}$, and $\{s_i\cdot q_i\}$.  
 \item If $\{s_i\}$  is a PR-Cauchy sequence then so is also $\{P_r(s_i)\}$, even uniformly on $r$, i.e., there is a binary PR function $M$ such that $|P_r(s_i)-P_r(s_j)|<2^{-k}$ for all $i,j\geq M(r,k)$.
 \item Every number in PRT is transcendental.
 \end{enumerate}
 \end{Lemma}

{\em Proof.} Items 1--3 are easy consequences of  Proposition 4.1 and Theorem 4.2 in \cite{zh07}. Item 4 (observed in the appendix of \cite{good61}) is obvious.
 \qed

\begin{Theorem}\label{prtrans2}
We have $\text{PRT}\sq\rp(\varkappa)$, hence every number from PRT may be included in some PRAS-field.
 \end{Theorem}

{\em Proof.} Let $\{s_i\}$  be a PR transcendental real; for the first assertion, we have to show that $s=\lim_is_i$ is in $\rp(\varkappa)$. By Theorem \ref{ext}, it suffices to show that  the sign of the value $P_r(s)$ may be found primitive recursively. Since $s$ is transcendental by Lemma \ref{prtrans1}(4), we get the dichotomy: $P_r(s)>0$ or $P_r(s)<0$. Therefore, it suffices to check that  $P_r(s)>0$ is a PR unary relation on $r$.

By Definition \ref{prtrans}, there are PR functions  $k,N$ such that $|P_r(s_n)|>2^{-k(r)}$ for every $n\geq N(r)$.  By Lemma \ref{prtrans1}, there is a binary PR function $M$ such that $|P_r(s_i)-P_r(s_j)|<2^{-k}$ for all $i,j\geq M(r,k)$. For the number $f(r)=max(N(r),M(r,k(r)+2))$ we then have $|P_r(s_{f(r)})|>2^{-k(r)}$ and $|P_r(s)-P_r(s_{f(r)})|<2^{-k(r)-1}$ (because $s=\lim_is_i$). Therefore, $P_r(s)>0$ iff $P_r(s_{f(r)})>0$. But $f$ is a PR function, so $P_r(s)>0$ is a PR  relation.

For the second assertion, let $s\in\text{PRT}$. Then $s\in\rp(\varkappa)$ by the first assertion. Since $\rat$ is a PRAS-field, by Theorem \ref{ext} so is also the ordered field $\rat(s)$ that contains $s$.
 \qed
  
\begin{Corollary}\label{prtrans3}
The ordered fields  $\rat(e)$, $\rat(\pi)$, $\widehat{\rat(e)}$, and $\widehat{\rat(\pi)}$ are PRAS-fields. 
 \end{Corollary}

{\em Proof.} In the appendix of \cite{good61} it is proved that  $e,\pi\in\text{PRT}$, hence $\rat(e)$ and $\rat(\pi)$ are PRAS-fields  by the proof of Theorem \ref{prtrans2}.   By Corollary \ref{rfinding1}(2),   $\widehat{\rat(e)}$ and $\widehat{\rat(\pi)}$ are also PRAS-fields.
 \qed

\begin{Remark}
Again, the results of this section are more intricate that their computable analogues in \cite{ss17a}. We illustrate this by the ordered field $\rat(e,\pi)$. By the general fact in \cite{ss17a} mentioned in the beginning of Section \ref{prreals}, $\rat(e,\pi)$ is a computable ordered field with splitting (the computable analogue of PRAS-fields). But we do not know whether it is a PRAS-field. The point is that it is open whether $e$ and $\pi$ are algebraically dependent (a long-standing open question in number theory). If yes, then $\rat(e,\pi)$ is a PRAS field by Corollary \ref{rfinding1}(ii). If not, a serious additional investigation is still needed to see whether $\pi\in\rp(\widehat{\alpha})$ where $\alpha$ is a PRAS-constructivization of $\rat(e)$.
 \end{Remark}

\section{PR linear algebra}\label{algebra}

Here we show PR computability of some problems of linear algebra frequently used in applications. 

As follows from an old result of F. Rellich (see e.g. \cite{zb04}),  the spectral problems discussed in this section are non-computable if we consider them, as is usually done in the books on linear algebra, for arbitrary real or complex matrices, and understand computability in the sense of A. Turing. This leads to computational instabilities when these problems are solved numerically, which was our original motivation to identify restricted computable versions of these problems. We will show that these problems are PR for matrices over any PRAS-field.

Recall that  {\em spectrum} of any complex matrix from $M\in M_n(\mathbb{C})$ is a sequence $\spec(M)=(\lambda_1,\ldots,\lambda_n)$ of all eigenvalues of $A$ (each eigenvalue occurs in the sequence several times, according to its multiplicity). The following fact immediately follows from Theorem \ref{rfinding1} and Proposition \ref{matrix}.

\begin{Proposition}\label{spec}
 Let $\mathbb{A}$ be a PRAS-field. Given $n$ and a symmetric matrix $M\in M_n(\widehat{\mathbb{A}})$, one can primitive recursively find a spectrum of $M$ uniformly on $n$.
\end{Proposition}

As is well known, all eigenvalues of any symmetric real matrix are real.  {\em Spectral decomposition} of such a matrix $A\in M_n(\mathbb{R})$ is a pair $((\lambda_1,\ldots,\lambda_n),(\mathbf{v}_1,\ldots,\mathbf{v}_n))$ where $\lambda_1\leq\cdots\leq\lambda_n$ is the non-decreasing spectrum of $A$  and $\mathbf{v}_1,\ldots,\mathbf{v}_n$ is a corresponding orthonormal basis of eigenvectors, i.e. $A\mathbf{v}_i=\lambda_i\mathbf{v}_i$ for $i=1,\ldots,n$. 

\begin{Proposition}\label{specmat}
  Let $\mathbb{A}$ be a PRAS-field. Given $n$ and a symmetric matrix $A\in M_n(\widehat{\mathbb{A}})$, one can primitive recursively find a spectral decomposition of $A$ uniformly on $n$.
\end{Proposition}

{\em Proof.} By Proposition \ref{spec}, we can primitive recursively compute  the increasing sequence $\mu_1<\cdots<\mu_m$ of all distinct roots of $ch_A$ and the corresponding multiplicities $r_1,\ldots,r_m$. Using the Gauss method, we then  primitive recursively  find, for each $j=1,\ldots,m$,  a basis $(\mathbf{w}^i_1,\ldots,\mathbf{w}^i_{r_j})$ for the eigenspace $\{\mathbf{x}\mid (\mu_j\cdot I_n-A)\mathbf{x}=\mathbf{0}\}$ corresponding to $\mu_j$. Applying the Gram-Schmidt orthonormalisation process $(\mathbf{w}_1,\ldots,\mathbf{w}_{r})\mapsto(\mathbf{v}_1,\ldots,\mathbf{v}_{r})$ which is (with simplified indices) given by the recurrent formulas
 $$\mathbf{v}_1=\mathbf{w}_1, \;\mathbf{v}_{k+1}=\mathbf{w}_{k+1}-\sum_{i=i}^k\frac{\mathbf{w}_{k+1}\cdot\mathbf{v}_{i}}{|\mathbf{v}_{i}|^2},$$
  we primitive recursively obtain an orthonormal basis $(\mathbf{v}^i_1,\ldots,\mathbf{v}^i_{r_j})$ for this eigenspace. Putting together the orthonormal bases for all $j$, we obtain a desired orthornormal basis $(\mathbf{v}_1,\ldots,\mathbf{v}_n)$ for the whole space.  Of course, the orthonormal basis of eigenvectors is not unique. It is only important that some such basis may be found primitive recursively. 
 \qed

The last proposition may be  extended to a natural class of complex matrices. A matrix $A\in M_n(\mathbb{C})$ is {\em normal} if $A\cdot A^*=A^*\cdot A$ where $A^*=(\bar{a}_{j,i})$ is the {\em conjugate matrix} for $A=(a_{i,j})$, and $A$ is {\em self-adjoint} if $A^*=A$. Note that every symmetric real matrix is self-adjoint, and every self-adjoint matrix is normal. It is known that for any eigenvalue $\mu$ of a normal matrix $A$ the multiplicity of $\mu$ coincides with the dimension of the corresponding eigenspace, and this eigenspace has a basis consisting of eigenvectors. The next extension of Proposition \ref{specmat} is thus proved by the same argument.
 
\begin{Proposition}\label{specmat1}
  Let $\mathbb{A}$ be a PRAS-field. Given $n$ and a normal matrix $A\in M_n(\overline{\mathbb{A}})$, one can primitive recursively find a spectral decomposition of $A$ uniformly on $n$.
\end{Proposition}

By a {\em matrix pencil} we mean a pair $(A,B)$ (often written in the form $\mu A-B$) of real non-degenerate symmetric matrices such that $A$ is positive definite (i.e., all of its eigenvalues are positive).
By {\em spectral decomposition} of such a pencil we mean a tuple
 $$((\lambda_1,\ldots,\lambda_n),(\mathbf{v}_1,\ldots,\mathbf{v}_n),(\mu_1,\ldots,\mu_n),(\mathbf{w}_1,\ldots,\mathbf{w}_n))$$
  such that $((\lambda_1,\ldots,\lambda_n),(\mathbf{v}_1,\ldots,\mathbf{v}_n))$ and $((\mu_1,\ldots,\mu_n),(\mathbf{w}_1,\ldots,\mathbf{w}_n))$ are spectral decompositions  of the symmetric matrices $A$  and $D^*L^*BLD$ respectively, where  $L$ is the matrix formed by vectors $\mathbf{v}_1,\ldots,\mathbf{v}_n$ written as columns, $L^*$ is the transposition of $L$, and $D=\operatorname{diag}\{\frac{1}{\sqrt{\lambda_1}}, \frac{1}{\sqrt{\lambda_2}}, \ldots, \frac{1}{\sqrt{\lambda_n}}\}$.
  
\begin{Proposition}\label{specpen}
   Let $\mathbb{A}$ be a PRAS-field. Given $n$ and a  matrix pencil $(A,B)$ with matrices $A,B$ in $ M_n(\widehat{\mathbb{A}})$, one can primitive recursively find a spectral decomposition of $(A,B)$ uniformly on $n$.
\end{Proposition}

{\em Proof.}  First we  find  a spectral decomposition $((\lambda_1,\ldots,\lambda_n),(\mathbf{v}_1,\ldots,\mathbf{v}_n))$ of $A$ using the algorithm of Proposition \ref{specmat}. Then we primitive recursively find  the matrix $D^*L^*BLD$. Applying the algorithm of Proposition \ref{specmat} to this matrix, we compute the remaining items $(\mu_1,\ldots,\mu_n),(\mathbf{w}_1,\ldots,\mathbf{w}_n)$. Note that $(\mu_1,\ldots,\mu_n)$ coincides with the spectrum of (in general, non-symmetric) matrix $A^{-1}B$. 
 \qed

An important normal form for square matrices over an algebraically closed field is the Jordan normal form. The next proposition gives a sufficient condition for PR-computability of this form similar to those given above. The proof, which is obtained by inspecting any standard  algorithm solving this problem written with all details (as e.g. the algorithm in Section 18 of \cite{gel}), is omitted.

\begin{Proposition}\label{jordan}
   Let $\mathbb{A}$ be a PRAS-field. Given $n$ and a  matrix  $M\in M_n(\overline{\mathbb{A}})$, one can primitive recursively and  uniformly on $n$ find a Jordan normal form $J\in M_n(\overline{\mathbb{A}})$ for $M$ and a non-degenerate matrix $C\in M_n(\overline{\mathbb{A}})$ with $M=C^{-1}JC$.
\end{Proposition}

\section{PR analysis}\label{anal}

 Computable analysis, developed  by many people,  was systematized in \cite{wei}. A central notion of computable analysis (due to A. Turing) is the notion of a computable (partial) function on the reals and on more complex spaces.
 In this section we  attempt  to develop a small part of PR analysis. Our approach is somewhat related to earlier works \cite{good61,gom11}. 
 
For any $n\geq0$, let  $\psi^{(n)}$ be the computable  numbering of  $n$-ary PR operators on the Baire space $\calN=\nat^\nat$ from the end of Section \ref{prel}. An example of unary PR operator is the function $q\to\tilde{q}$ defined in Section \ref{prreals}. This operator is clearly a retraction of $\calN$ onto $\cau$, i.e. $\tilde{q}\in\cau$ for each $q\in\calN$ and $\tilde{q}=q$ for  $q\in\cau$. Recall that, by the definition, $q\in\cau$ iff $\varkappa\circ q$ is fast Cauchy. Another example is the function $q\to q'$ defined at the end of Section \ref{prreals}. Note that every PR operator on $\calN$ is total and the class of such operators is closed under composition.

 We proceed with defining PR functions on the reals.
Similar to the ideas of computable analysis, we can use a suitable surjection $\gamma:\calN\to\IR$ to transfer primitive recursiveness on $\calN$ to that on $\IR$ (and then, may be, to more complex spaces). Namely, we define $\gamma(p)=\lim_n\varkappa(\tilde{p}(n))$  and call this $\gamma$ the {\em Cauchy representation} of $\IR$. Note that $\gamma$ is an admissible representation in the sense of \cite{wei}, and it is computably equivalent to the Cauchy representation of $\IR$ defined in \cite{wei}.

\begin{Definition}\label{prrealfunc}
A partial function $f:\sq\IR^{n}\to\IR$ is called PR if there is a PR function $g:\calN^{n}\to\calN$ (called a $\gamma$-realizer of $f$) such that $f(\gamma(p_1),\ldots,\gamma(p_n))=\gamma(g(p_1,\ldots,p_n))$ for all $p_1,\ldots,p_n\in\calN$.
\end{Definition}

Although this definition looks completely similar to the definition of a computable partial function on the reals, there is an important difference: while in the computable case the computable realizer may be partial, the PR-realizers are always total, as any PR operator on $\calN$.  
Clearly, the PR functions on the reals are computable, and in fact they form a very restricted subclass of the computable functions. Nevertheless, many practically important functions are PR. We illustrate similarities and differences between computable and PR functions by providing some examples.

\begin{Proposition}\label{samples}
\begin{enumerate}\itemsep-1mm
 \item The functions $+,\cdot,-$ on $\IR$ (and on $\co$) are PR.
 \item For every polynomial  $f\in\rp[x_1,\ldots,x_n]$, the corresponding evaluation function $f_p:\IR^n\to\IR$ is PR.
 \item The reciprocal partial  function $x^{-1}$ on $\IR$ (and on $\co$) is computable but not PR (cf. Remark \ref{pi0}).
 \item The class of PR functions (and the class of PR partial functions) on $\IR$ is closed under composition.
 \item The 0-ary total functions on $\IR$ coincide with the PR reals.
 \item  Let $f:\sq\IR^{n}\to\IR$ be a PR partial function on the reals. Then its restriction $f|_{\rp}$ to $\rp$ is $\tau$-PR. 
 \end{enumerate}
 \end{Proposition}

{\em Proof.} 1. This is proved by essentially the same argument as that in the proof of Proposition \ref{pi}(1).

2. Follows from item 1.

3. The computability of the reciprocal partial  function is well known (with a computable partial realizer). Suppose for a contradiction that it is PR and let $g$ be its PR realizer (which is automatically total). Then $g$  is computable, hence also continuous. Let $q=\{2^{-i}\}$ and let $r=\{a_i\}$ be the unique sequence of rationals with $g(\varkappa^{-1}\circ q)=\varkappa^{-1}\circ r$. Since $g$ is continuous, for a suitable $n$ we have: for any fast Cauchy sequence $q'=\{q_i\}$ of rationals with $q_0=2^{-0},\ldots,q_n=2^{-n}$, the corresponding sequence $r'=\{a'_i\}$ of rationals with $g(\varkappa^{-1}\circ q')=\varkappa^{-1}\circ r'$, satisfies $a'_0=a_0$. Let $x=\lim q'$ and $y=\lim(\varkappa\circ\tilde{p})$ where $p=\varkappa^{-1}\circ r'$. Note that the first element of the sequence $\varkappa\circ\tilde{p}$ is $a_0$, and that $x^{-1}=y$ whenever $x\not=0$ (because $g$ is a realizer for $^{-1}$). Since $q'$ and $\varkappa\circ\tilde{p}$ are fast Cauchy, we have $x\in[2^{-n}-2^{-n+1},2^{-n}+2^{-n+1}]$, $y\in[a_0-2,a_0+2]$.

Let now $m>n$ be any integer with $2^m>a_0+2$, and define $q'=\{q_i\}$ as follows: $q_0=2^{-0},\ldots,q_n=2^{-n}$, and $q_{n+i+1}=2^{-m}$. Then $x=2^{-m}$, hence $y=x^{-1}=2^m\not\in[a_0-2,a_0+2]$. A contradiction.  

4. The assertion follows from the corresponding fact for realizers which are PR operators on $\calN$. 

5. Obvious.

6. Let $g$ be a PR realizer for $f$, then $g=\mathbf{u}$ for some $u(v_1,\ldots,v_n)\in T_n$ (we use the notation and facts from the end of Section \ref{prel}). It suffices to show that the PR function $h(k_1,\ldots,k_n)=N(u(t_{k_1},\ldots t_{k_n}))$ satisfies $f(\tau(k_1),\ldots,\tau(k_n))=\tau(h(k_1,\ldots,k_n))$ whenever $f(\tau(k_1),\dots,\tau(k_n))$ is defined. Indeed, we have:
\begin{eqnarray*}
 f(\tau(k_1),\dots,\tau(k_n))=f(\gamma(\psi(k_1)),\ldots,\gamma(\psi(k_n))=\gamma(g(\psi(k_1),\ldots,\psi(k_n))\\=g(\bft_{k_1},\ldots,\bft_{k_n})=\gamma(\psi(h(k_1,\ldots,k_n)))=\tau(h(k_1,\ldots,k_n)).
 \end{eqnarray*}

 \qed

\begin{Remark}\label{samples1}
PR analysis is much more intricate than computable analysis. For instance, the computable version of the basic notion of ``elementary function'' is straightforward but the corresponding ``correct'' PR analogue is far from obvious, as demonstrated in particular by the facts about the reciprocal function. Nevertheless, we guess that many important concrete functions are PR. For instance, the results in the appendix to \cite{good61} suggest that this is  the case for the functions $sin,cos$ though some additional work is needed to adjust the arguments in \cite{good61} to our context. 
\end{Remark}

An important notion of computable analysis is the notion of a computable metric space. Having in mind applications in the next section, we propose the following PR versions of this notion.

\begin{Definition}\label{prmetric}
\begin{enumerate}
 \item By a PR metric space we mean a triple $(M,d,\nu)$ where $(M,d)$ is a metric space, $\nu:\nat\to M$ is a numbering of a dense subset $M_0\sq M$, and there is a ternary PR sequence $\{q_{m,n,i}\}$ of rationals such that, for all $m,n$, $\{q_{m,n,i}\}_i$ is fast Cauchy $d(\nu_m,\nu_n)=\lim_iq_{m,n,i}$.
 \item For $\alpha\in\pr(\IR)$, by an $\alpha$-PR metric space we mean a triple $(M,d,\nu)$ as above such that $d(\nu_m,\nu_n)=\widehat{\alpha}_{f(m,n)}$, for a suitable  binary PR function $f$.
 \end{enumerate}
  \end{Definition}

With any triple $(M,d,\nu)$ as above we can in the usual way associate the so called Cauchy representation of $M$, i.e., the partial function $\nu^\ast$ from $\mathcal{N}$ onto $M$ as
follows: $\nu^\ast(p)=x$ iff the sequence $\{\nu_{p(n)}\}$ is
fast Cauchy and converges to $x$.
 For every  $\alpha$-PR metric space  $(M,d,\nu)$,  the Cauchy representation is constructed more regularly. Namely, let the set $\cau=\cau_\nu\sq\calN$ and the function $p\mapsto\tilde{p}=\tilde{p}_\nu$ on $\calN$ be defined in the same way as the corresponding objects in Section \ref{prreals} but with $\nu$ in place of $\varkappa$ and $d$ in place of the usual distance function on the reals. Then again $p\mapsto\tilde{p}$ is a PR retraction from $\calN$ onto $\cau$. We set $\nu^*(p)=\lim(\nu\circ\tilde{p})$ whenever the limit exists. 

\begin{Proposition}\label{prmetric1}
\begin{enumerate}\itemsep-1mm
 \item Any $\varkappa$-PR metric space is an $\alpha$-PR metric space for every $\alpha\in\pr(\IR)$.
 \item If $(M,d,\nu)$  is an $\alpha$-PR metric space and $(M,d)$ is complete then $\nu^*$ is total.
 \item For every infinite $\alpha$-PR metric space  $(M,d,\nu)$ there is an injective numbering $\nu'\equiv_{PR}\nu$, so $(M,d,\nu')$ is an $\alpha$-PR metric space and $\nu^*\equiv_{PR}\nu'^*$.
 \item Every $\alpha$-PR metric space is a PR metric space. 
 \end{enumerate}
 \end{Proposition}

{\em Proof.} 1. This follows from $\varkappa\leq_{PR}\alpha$.

2. Obvious by the definition of Cauchy representation.

3. The numbering $\nu'$ is constructed from $\nu$ in the same way as the numbering $\gamma$ is constructed from $\beta$ in the proof of Proposition \ref{fpr}. The properties of $\nu'$ are obvious. Note that, say, the reduction $\mu\leq_{PR}\nu$ between partial representations means that there is a unary PR operator $f$ on $\calN$ such that $\mu_p=\nu_{f(p)}$ for every $p\in dom(\mu)$.

4. Let $f$ be  a binary PR function $f$ such that $d(\nu_m,\nu_n)=\widehat{\alpha}_{f(m,n)}$.  Proposition \ref{rpalpha}(4) yields the sequence $\{q_{m,n,i}\}$ witnessing that $(M,d,\nu)$ is a PR metric space.
 \qed

Definition \ref{prmetric} naturally extends to the definition of a PR function between PR metric spaces:

\begin{Definition}\label{prmetric0}
 Let  $(M,d,\nu)$ and  $(M_1,d_1,\nu_1)$ be PR metric spaces. A  partial function $f:\sq M\to M_1$ is PR if it has a PR realizer w.r.t. $\nu^*,\nu_1^*$, i.e., there is a unary PR operator $g$ on $\calN$ such that $\nu_1^*(g(p))=f(\nu^*(p))$ for every $p\in dom(\nu^*)$.
\end{Definition}

For readers' convenience, we give  examples of PR metric spaces (and normed spaces) and PR functions between them relevant to the next section; for more detailed definitions of these spaces see e.g. \cite{ss17} and references therein. For any $n\geq 1$, the Euclidean vector space  ${\mathbb R}^n$ carries the
sup-norm $||x||_\infty=\operatorname{max}\{|x_i|\}$  and the Euclidean norm
$||x||_2=\sqrt{\sum x_i^2}$; we denote the corresponding metrics
by $d_\infty$ and $d_2$, respectively.
Define the function $\varkappa^n:\mathbb{N}\to{\mathbb R}^n$ by
$\varkappa^n(\langle
k_1,\ldots,k_n\rangle)=(\varkappa(k_1),\cdots,\varkappa(k_n))$.

There are several natural subspaces and variations of the Euclidean spaces, in particular the space $S\subseteq{\mathbb R}^{n\times n}$ of symmetric real matrices, the space $S_{+}$  of symmetric real positively definite matrices, and the $m$-dimensional unitary cube $Q=[0,1]^m$. They have numberings of dense subsets induced by the corresponding numbering $\nu\in\{\varkappa^{n^2},\varkappa^n\}$, using the easy fact that the sets $\nu^{-1}(S)$, $\nu^{-1}(S_+)$, and $\nu^{-1}(Q)$ are PR.

Important modifications of the Euclidean spaces are the spaces of grid functions. Consider, for any positive integer $N$, the uniform rectangular  grid $G_N$ on $Q=[0,1]^m$ defined
by the  points $$\left(\frac{i_1-\frac{1}{2}}{2^N},\frac{i_2-\frac{1}{2}}{2^N},\ldots,\frac{i_m-\frac{1}{2}}{2^N}\right)$$ where $1\leq
i_1,i_2, \ldots, i_m\leq2^N$. Let
$h=1/2^N$ be the corresponding spatial grid step and  $\tau$ be a time step. Denote  $G_N^{\tau}=G_N\times\{l\tau\}_{l=1}^L$, where $L$ is the number of the time steps. We will consider  the  grid norms
 $$||g^{(h)}||_s=\operatorname{max}_{x\in G_N}|g^{(h)}(x)|, \;
 ||g^{(h)}||^2_{L_2}=h^m\sum_{x\in G_N}\langle g^{(h)}(x),g^{(h)}(x)\rangle$$
 on the grid functions $g^{(h)}:G_N\to\mathbb{R}^n$ and the $sL_2$-norm
$$||v^{(h)}||_{sL_2}=\operatorname{max}_{t\in\{l\tau\}_{l=1}^M}h^m\sum_{x\in G_N}\langle v^{(h)}(t,x),v^{(h)}(t,x)\rangle$$
 on the grid functions $v^{(h)}(t,x):G^\tau_N\to\mathbb{R}^n$. The numberings of dense sets (formed by the rational functions on the grids) are induced by the numbering $\varkappa$.

We  work with several functional spaces most of which are
subsets of the set $C({\mathbb R}^m,{\mathbb R}^n)\simeq
C({\mathbb R}^m,{\mathbb R})^n$ of integrable continuous functions
$\varphi:{\mathbb R}^m\rightarrow{\mathbb R}^n$ equipped with  the
$L_2$-norm. In particular, we deal with the space $C(Q,{\mathbb
R}^n)\simeq C(Q,{\mathbb R})^n$ (resp. $C^k(Q,{\mathbb R}^n)$) of
continuous (resp. $k$-time  continuously differentiable) functions
$\varphi:Q\rightarrow{\mathbb R}^n$ equipped with the $L_2$-norm
 $$||\varphi||_{L_2}=\left(\int_Q|\varphi(x)|^2dx)\right)^{\frac{1}{2}},\; |\varphi(x)|^2=
\langle\varphi,\varphi\rangle=\sum\limits_{i=1}^n\varphi^2_i(x).$$
 We will also use the  sup-norm $$||\varphi||_s=\sup_{x\in
Q}|\varphi(x)|,\ ||f||_s=\sup_{(t,x)\in[0,T]\times Q}|f(t,x)|$$ on
$C(Q,\mathbb R^n)$ or $C([0,T]\times Q,\mathbb R^n)$ and the
$sL_2$-norm
 $$||u||_{sL_2}=\sup_{0\leq t_0\leq
 T}\sqrt{\int_Q|u(t_0,x)|^2dx}$$
  on
$C([0,T]\times Q,{\mathbb R^n})$ where $T>0$.  Whenever we want to
emphasize the norm we use notation like $C_{L_2}(Q,{\mathbb
R}^n)$, $C_s(Q,{\mathbb R}^n)$ or $C_{sL_2}([0,T]\times Q,{\mathbb
R}^n)$.

Associate with any rational grid function $f_N:G_N\rightarrow{\mathbb Q}$ the
continuous extension $\tilde{f}_N:Q\rightarrow{\mathbb R}$ of $f$
obtained by piecewise-linear interpolation on each coordinate, and similarly for grid functions $G^\tau_N\to\rat$. Such
interpolations are known  as {\it multilinear interpolations}.
Note that the restriction of $\tilde{f}_N$ to any grid cell is a
polynomial of degree $m$. The
extensions $\tilde{f}_N$ induce a countable dense set in
$C(Q,{\mathbb R}^n)$ (or $C([0,T]\times Q,{\mathbb R^n})$) with any
of the three norms induced by $\varkappa$ and natural numberings of the grids $G_N$ for all $N\geq1$ and  of the grids $G^\tau_N$ for all $N\geq1$ and rational $\tau$.

Along with the mentioned norms,  their
$A$-modifications, for a given matrix $A$, are useful. In particular, the
$A$-modification of the $L_2$-norms is defined by
 $$
||\varphi||_{A,L_2}=\sqrt{\int_Q\langle A\varphi,\varphi\rangle dx}
 $$
 while the $A$-modification of the $sL_2$-norms is defined by
 $$
||u||_{A,sL_2}=\sup_{0\leq t_0\leq
 T}\sqrt{\int_Q\langle Au(t_0,x),u(t_0,x)\rangle dx},
 $$
 and in a similar way for the grid norms.

\begin{Proposition}\label{prmetric2}
\begin{enumerate}\itemsep-1mm
 \item The spaces   $({\mathbb R}^n,d_\infty,\varkappa^n)$ and $({\mathbb R}^n,d_2,\varkappa^n)$ are $\varkappa$-PR metric spaces uniformly on $n$.
 \item The spaces $S,S_+,Q$ and the spaces of grid functions  are $\varkappa$-PR metric spaces uniformly on their dimensions.
 \item For any $n\geq 1$, the spaces
$C_s(Q,{\mathbb R}^n)$, $(C_{L_2}(Q,{\mathbb
R}^n)$,  and $C_{sL_2}([0,T]\times Q,{\mathbb R}^n)$ are $\varkappa$-PR metric spaces uniformly on $n$.
 \item  Let $\alpha\in\prs(\IR)$ and let $A$ be a symmetric matrix in $M_n(\widehat{\mathbb{A}})$. Then  the spaces from  (3), equipped with the $A$-modified metrics, are $\alpha$-PR metric spaces uniformly on $n$.
 \item The restriction
$\varphi\mapsto\varphi|_G$ is a PR function from
$C_s(Q,{\mathbb R}^n)$ to $({\mathbb R}^n)^G$.
 \item The operator $f\mapsto\tilde{f}$ is a
PR function from $(({\mathbb R}^n)^G)_s$ to
$C_{L_2}(Q,{\mathbb R}^n)$
 \end{enumerate}
 \end{Proposition}

{\em Proof.} Items (1,2) are obvious from the definition.

3. The proof is a straightforward calculation which shows that the distance between elements of the dense subsets is an algebraic real which is uniformly primitive recursively computable. Item (6) is considered similarly, only now calculations are made within the PRAS-field $\widehat{\mathbb{A}}$.

Items (4,5) follow from the definitions and well-known properties of the multilinear interpolations.
 \qed

\section{PR solutions of PDE}\label{prpde}

Here we apply the above-developed theory to investigating the question when the solution operators for hyperbolic symmetric systems of PDE are PR. This complements the results in \cite{ss17a} and \cite{ss18} where computability and bit complexity of such operators were examined.

For simplicity we discuss here only the Cauchy initial value problem (the boundary value problems are considered similarly) stated as follows:
\begin{equation} \label{sist_1}
\begin{cases} A\frac{\partial
{\bf u}}{\partial t}+\sum\limits_{i=1}^mB_i\frac{\partial {\bf u}}
{\partial x_i}=f(t,x_1,\ldots,x_m),\ t\geq 0,\\
{\bf u}|_{t=0}=\varphi(x_1,\ldots,x_m),
\end{cases}
\end{equation}
where $A=A^\ast>0$ and $B_i=B_i^\ast$ are non-degenerate symmetric
 $n\times n$-matrices, $t\geq0$, $x=(x_1,\ldots,x_m)\in
Q=[0,1]^m$, $\varphi:Q\rightarrow{\mathbb R}^n$, $f:[0,+\infty)\times Q\rightharpoonup{\mathbb R}^n$ and ${\bf
u}:[0,+\infty)\times Q\rightharpoonup{\mathbb R}^n$ is a partial
function acting on the domain $H$ of existence and uniqueness of
the Cauchy problem \eqref{sist_1}.  The set $H$ is known to be (see e.g.
\cite{ss18} for references and additional information) the intersection of  semi-spaces
 $$
t\geq0,\;x_i-\mu^{(i)}_{\rm
max}t\geq0,\;x_i-1-\mu^{(i)}_{\rm min}t\leq0\;(i=1,\ldots,m)
 $$
of ${\mathbb R}^{m+1}$ where $\mu^{(i)}_{\rm
min},\mu^{(i)}_{\rm
max}$ are  the minimum and maximum of the eigenvalues of  $A^{-1}B_i$.

Below we formulate sufficient conditions for PR-computability of the solution operator for \eqref{sist_1} by putting some restrictions on matrices $A,B_1,\ldots,B_m$ and functions $\varphi,f$. Matrix coefficients are usually in a real closed PRAS-field $\mathbb{A}$. The results below are PR analogues of the corresponding results in \cite{ss17,ss18}. Most of the technical details are the same as in those papers, so we only give short proof sketches pointing to the  new details.  

We start with formulating two auxiliary facts.
The next immediate corollary of Propositions \ref{specmat} and \ref{specpen} shows that we can primitive recursively compute $H$. Our algorithms for solving the Cauchy problem are for technical reasons presented only for the case when $H$ satisfies the condition $\mu^{(i)}_{\rm
min}<0<\mu^{(i)}_{\rm max}$ for all $i=1,\ldots,m$ (which implies that $H$ is compact); this condition
often holds for natural physical systems. 

\begin{Proposition}\label{cpH}
Let $\mathbb{A}$ be a PRAS-field. Given  $m,n\geq1$ and $A,B_1\ldots,B_m\in M_n(\widehat{\mathbb{A}})$ as in \eqref{sist_1}, one can compute $(\mu^{(1)}_{\max},\ldots,\mu^{(m)}_{\max},
\mu^{(1)}_{\min},\ldots,\mu^{(m)}_{\min})$
 and check the condition $\mu^{(i)}_{\rm
min}<0<\mu^{(i)}_{\rm max}$ for all $i=1,\ldots,m$ primitive recursively uniformly on $m,n$. Thus, the algorithm finds the domain $H$ satisfying the condition above, or reports on the absence of such a domain.
 \end{Proposition}

We also need another immediate corollary of  Propositions \ref{specmat} and \ref{specpen} about the spectral decomposition $((\lambda_1,\ldots,\lambda_n),(\mathbf{v}_1,\ldots,\mathbf{v}_n))$     of the matrix $A$ as in \eqref{sist_1}. Let $\lambda_{max}$, $\lambda_{min}$ be respectively the maximum and minimum of $\lambda_1,\ldots,\lambda_n$. Let $L$ be the orthonormal matrix formed by vectors $\mathbf{v}_1,\ldots,\mathbf{v}_n$ written in columns, so
 $L^*AL=\Lambda={\rm diag}\{\lambda_1,\lambda_2,\ldots,\lambda_n\}$, and let
$D=\Lambda^{-\frac{1}{2}}$. 
For each $i=1,\ldots,m$, let $((\mu^{(i)}_1,\ldots,\mu^{(i)}_n),(\mathbf{w}^{i}_1,\ldots,\mathbf{w}^{i}_n))$ be the spectral decomposition  of the symmetric matrix $D^*L^*B_iLD$. Let $\mu^{(i)}_{max}$, $\mu^{(i)}_{min}$ be respectively the maximum and minimum of $\mu^{(i)}_1,\ldots,\mu^{(i)}_n$. Let $M_i={\rm
 diag}\{\mu^{(i)}_1,\ldots,\mu^{(i)}_n)\}$ and  $K_i$ be the orthonormal matrix formed by vectors $\mathbf{w}^{i}_1,\ldots,\mathbf{w}^{i}_n$ written in columns, so $K_i^*D^*L^*B_iLDK_i=M_i$. Let $T_i=LDK_i$ for each $i=1,\ldots,m$. 

\begin{Proposition}\label{schemedata}
Let $\mathbb{A}$ be a PRAS-field. Given $m,n\geq 1$ and $A,B_1\ldots,B_m\in M_n(\widehat{\mathbb{A}})$ as in \eqref{sist_1}, one can primitive recursively uniformly on $m,n$ compute the quantities $A^{-1}$, $T_i$, $T_i^{-1},\lambda_{max}$, $\lambda_{min},\mu^{(i)}_{max},\mu^{(i)}_{min}$, $\mu^{(i)}_k$($i=1,\ldots,m,k=1,\ldots, n$) specified above. 
\end{Proposition} 
 
Now we are able to formulate our results about PR-computability of the Cauchy problem. The last proposition contains all information used in the Godunov difference scheme which in the central part of our algorithm. All  details of the algorithm are given in \cite{ss18}. The next result is a PR-version of Theorem 5.2 in \cite{ss17a} (though  Theorem 5.2 was about the boundary-value problem). Moreover, for notation simplicity we stick to the case $f\equiv 0$, as in \cite{ss18}.

\begin{Theorem}\label{pras-coef}
Let $\mathbb{A}$ be a PRAS-field and let $M,  p\geq2$ be integers. Then the solution operator
 $(A,B_1,\ldots,B_m,\varphi)\mapsto{\mathbf u}$ for \eqref{sist_1} is a PR-computable function (uniformly on $m,n$)  from  $S_{+}\times S^{m}\times 
C_s^{p+1}(Q,{\mathbb R}^n)$ to $C_{sL_2}^p(H,{\mathbb R}^n)$ where $S$ and $ S_{+}$ are respectively the sets of all symmetric and symmetric positively definite matrices from $M_n(\widehat{\mathbb{A}})$, $
||\frac{\partial
\varphi}{\partial x_i}||_{s}\leq M$ and $\ ||\frac{\partial^2
\varphi}{\partial x_i\partial x_j}||_{s}\leq M$ for $ i,j=1,2,\ldots,m.$
\end{Theorem}

{\em Proof.} We first make precise computations as in Propositions \ref{cpH} and \ref{schemedata}. 
With these quantities at hand, we repeat  computations with the Godunov scheme as in Section 5.2. of \cite{ss17a}. Namely, we primitive recursively transform any given sequence $\{\varphi_k\}$ of grid functions $\varphi_k:G_{i_k}\rightarrow{\mathbb Q}^n$  such that their multilinear
interpolations $\{\tilde{\varphi}_k\}$ form a fast Cauchy sequence
converging in $C_s(Q,{\mathbb R}^n)$ to $\varphi$, to a fast Cauchy sequence
converging in $C_{sL_2}^p(H,{\mathbb R}^n)$ to  the solution of the Cauchy problem. For any $k$ we first compute the time step $\tau_k$ as in Lemma 4.4 of \cite{ss17}; this computation is PR. Then we compute the grid function ${\mathbf\upsilon}_k:G^\tau_{i_k}\rightarrow{\mathbb A}^n$ by the algorithm of Godunov's scheme; these computations are performed precisely within the field $\widehat{\mathbb{A}}$ and they are PR since  the involved metric spaces are $\alpha$-PR (see Propositions \ref{prmetric2} and \ref{prmetric1}(1)). 

In this way, we primitive recursively transform any given sequence $\{\varphi_k\}$ to the sequence $\{\upsilon_k\}$. As shown in Sections 5.2 and 5.3 of \cite{ss17a}, the corresponding sequence $\{\tilde{\upsilon}_k\}$ of interpolations  satisfies   $||\tilde{\upsilon}_k-{\bf u}||_{sL_2}\leq c\cdot 2^{-k}$ where $c$ is a constant depending only on $A,B_1,\ldots,B_n,M$. Moreover,  the estimates in Sections 5.2 and 5.3 of \cite{ss17} show that $c$ is also computed primitive recursively. Let $m$ satisfy $c<2^m$, then the $m$-shift of $\{\tilde{\upsilon}_k\}$ is the desired fast Cauchy PR sequence of approximations to the solution of \eqref{sist_1}.
 \qed

For fixed $A,B_1\ldots,B_m\in M_n(\widehat{\mathbb{A}})$, Theorem \ref{pras-coef} of course implies PR-computability of the solution operator $\varphi\mapsto{\mathbf u}$ for \eqref{sist_1}. The next result (which is a PR-version of Theorem 5.1 in  \cite{ss17a}) shows that the assumption  $A,B_1\ldots,B_m\in M_n(\widehat{\mathbb{A}})$ may be weakened to $A,B_1\ldots,B_m\in \rp$. 

\begin{Theorem}\label{fixed-coef}
Let $M,  p\geq2$ be integers and $A,B_1,\ldots,B_m\in M_n(\rp)$ be fixed matrices satisfying the conditions in \eqref{sist_1} such that $\mu^{(i)}_{\rm
min}<0<\mu^{(i)}_{\rm max}$ for all $i=1,\ldots,m$. Then the solution operator
 $\varphi\mapsto{\mathbf u}$ for \eqref{sist_1} is a PR-computable function (uniformly on $m,n$) from $
C_s^{p+1}(Q,{\mathbb R}^n)$ to $C_{sL_2}^p(H,{\mathbb R}^n)$,whenever $\varphi$ satisfies the conditions $
||\frac{\partial
\varphi}{\partial x_i}||_{s}\leq M$ and $\ ||\frac{\partial^2
\varphi}{\partial x_i\partial x_j}||_{s}\leq M$ for $ i,j=1,2,\ldots,m.$
\end{Theorem}

{\em Proof.} All the quantities from Propositions \ref{cpH} and \ref{schemedata} are fixed elements of $\rp$ because this field is real closed by Proposition \ref{pi}(4). With these data at hand, all computations in the Godunov scheme are made within $\rp$ using only the operations $+,\cdot,-$, hence these computations are $\tau$-PR by Proposition \ref{pi}(1) where $\tau$ is the numbering of $\rp$. Thus, the grid functions $\upsilon_k$ take now values in $\rp$ uniformly on $k$. In fact, we can work directly with the corresponding fast Cauchy PR sequences of rationals, and the family of all such sequences is uniformly PR, hence we can primitive recursively compute arbitrarily precise rational approximations to the grid functions $\upsilon_k$.
 \qed

\begin{Remark}
Let $n_A,n_1,\ldots,n_m$ be  the cardinalities of spectra of $A$ and of the matrix pencils $\mu A- B_1,\ldots,\mu A-B_m$, resp. Theorem 5.3 in \cite{ss17a} (formulated there also for $f=0$) states that the operator
 $(A,B_1,\ldots,B_m,n_A,n_1,\ldots,n_m,\varphi)\mapsto{\mathbf u}$
  sending any  sequence
$A,B_1,\ldots,B_m$ of symmetric real matrices  (with some restrictions similar to those in Theorem \ref{pras-coef}) such that the matrix pencils $\mu A- B_i$ have no zero eigenvalues, to the solution of \eqref{sist_1},
 is a computable
partial function from the space $S_{+}\times S^{m}\times \mathbb{N}^{m+1}\times
C_s^{p+1}(Q,{\mathbb R}^n)$ to $C_{sL_2}^p(H,{\mathbb R}^n)$. The PR-version of this theorem would give just its strengthening to the claim that the operator is PR. Unfortunately, the proof in  \cite{ss17} does not straightforwardly yield  this strengthening  because it first computes ``good enough'' rational approximations to $A,B_1,\ldots,B_m$ but this computation makes use of the unbounded search.
\end{Remark}

We conclude this section with the following PR-version of Theorem 2 in \cite{ss18}. Note that formulation is broader than in \cite{ss18} because now the algorithm is uniform on $m,n,a$ and works not only with algebraic numbers. 

\begin{Theorem}\label{cpq}
Let $\mathbb{A}$ be a PRAS-field. Given integers $m,n,a\geq 1$, matrices $A,B_1\ldots,B_m\in M_n(\widehat{\mathbb{A}})$, and rational functions $\varphi_1\ldots,\varphi_n\in \widehat{\mathbb{A}}(x_1\ldots,x_m), \;f_1\ldots,f_n\in \widehat{\mathbb{A}}(t,x_1\ldots,x_m)$ as in \eqref{sist_1}, one can primitive recursively uniformly on $m,n,a$ compute  a rational $T>0$ with $H\subseteq[0,T]\times Q$, a spatial rational grid step $h$ dividing $1$, a time grid step $\tau$ dividing $T$ and an $h,\tau$-grid function $v:G_N^{\tau}\to\widehat{\mathbb{A}}$ such that $||{\bf u}-\widetilde{\upsilon\mid_H}||_{sL_2}<a^{-1}$, where $\widetilde{\upsilon\mid_H}$ is the multilinear interpolation of the restriction of the grid function $\upsilon$ to $H$.
\end{Theorem}

{\em Proof.} The proof is an easy modification of the proof of Theorem 2 in \cite{ss18}. First we use  Propositions \ref{cpH} and \ref{schemedata} to primitive recursively compute input data for the difference scheme algorithm. Then we use  the algorithm of Section 4.1 in \cite{ss18} to compute the space and time grid steps $h,\tau$ which guarantee the estimate $||{\bf u}-\widetilde{\upsilon\mid_H}||_{sL_2}<a^{-1}$ (for fixed parameters this algorithm works in polynomial time but uniformity requires a bit higher complexity). Then we use the algorithm of Godunov's scheme which is written in detail in \cite{ss18} as an explicit sequence of matrix computation which are precise computations within the field $\widehat{\mathbb{A}}$. The results of the current paper show that these computations are PR. Note that the resulting grid function $\upsilon:G_N^\tau\to \mathbb{A}^n$ takes values in $\mathbb{A}$. If needed, Proposition \ref{rpalpha}(4) may be used to primitive recursively compute a rational-valued grid function with the same precision estimate.
 \qed

\section{Conclusion} \label{open}

Papers \cite{ss17,ss17a,ss18}, and also this paper, is a mix of symbolic algorithms (which aim to find  precise solutions), and  approximate algorithms (which aim to find  ``good enough'' approximations to precise solutions). The symbolic algorithms  implemented e.g. in computer algebra systems correspond well to computations on discrete structures (with mathematical foundations in the classical computability and complexity theory). The approximate algorithms included in many numerical mathematics packages  correspond well to computations on continuous structures (with mathematical foundations in the field of computability and complexity in analysis).

We hope that the present paper demonstrates that PR computations is a natural next step in the investigation of this interaction because it provides a natural borderline between problems in  algebra and analysis computable in principle and feasible problems. Although  PR functions were thoroughly investigated in computability theory and proof theory, their study in computable structure theory and computable analysis seems still in the very beginning. 

The fast progress of computation theory in the last decades made  informal descriptions of several  algorithms in  standard textbooks in algebra and analysis insufficient and sometimes even incorrect. They typically remain correct when interpreted in countable discrete structures like those in Section \ref{prings}, though a finer distinction between general computability and feasible computability is desirable. Some algorithms interpreted in continuous structures (like the real or complex numbers) become even incorrect (see examples in Section \ref{algebra}); it seems desirable to add corresponding comments (which mention the computable analysis approach) in the next editions of such textbooks.

\medskip

{\bf Acknowledgements.} 
The  authors thank Pavel Alaev, Sergey Goncharov, Valentina Harizanov, Peter Hertling, Iskander Kalimullin, Julia Knight, Russell Miller, Andrey Morozov, and Xizhong Zheng for   useful discussions and bibliographical hints.

\end{document}